\documentclass[journal,twocolumn]{IEEEtran}

\usepackage[final]{graphicx}

\usepackage{amsmath,epsfig,amssymb,verbatim,amsopn,cite,subfigure,multirow,amsthm}
\usepackage{balance}
\usepackage[usenames,dvipsnames]{color}
\usepackage[all]{xy}  
\usepackage{url}
\usepackage{amsfonts}
\usepackage{amssymb}
\usepackage{epsfig}
\usepackage{bm}
\usepackage{epsfig}
\usepackage{amsmath,amssymb, amstext}
\usepackage{graphicx}
\usepackage{epstopdf}
\usepackage{algorithm}
\usepackage{algorithmicx}
\usepackage{tabularx,booktabs}
\usepackage{url}
\usepackage{multirow}
\usepackage{array}
\usepackage{makecell}
\usepackage{balance}
\usepackage{algorithm}
\usepackage{algpseudocode}
\usepackage{algcompatible}
\usepackage{booktabs,arydshln}
\usepackage{nicematrix}
\usepackage{subfig}
\usepackage{caption}
\usepackage{algorithm}
\usepackage{algpseudocode}
\usepackage{algcompatible}

\usepackage{accents}
\newlength{\dhatheight}
\newcommand{\doublehat}[1]{%
    \settoheight{\dhatheight}{\ensuremath{\hat{#1}}}%
    \addtolength{\dhatheight}{-0.35ex}%
    \hat{\vphantom{\rule{1pt}{\dhatheight}}%
    \smash{\hat{#1}}}}

\DeclareMathOperator*{\argmax}{arg\,max}

\begin{document}

\title{Spectral-Efficient LoRa with Low Complexity Detection
}

\author{Alireza Maleki, Ebrahim Bedeer, and Robert Barton
\thanks{Alireza Maleki and Ebrahim Bedeer are with the Department of Electrical and Computer Engineering, University of Saskatchewan, Saskatoon, Canada S7N5A9. Emails: \{alireza.maleki and e.bedeer\}@usask.ca.}
\thanks{R. Barton is with Cisco Systems Inc. Email: robbarto@cisco.com.}
\thanks{This work was supported by the NSERC/Cisco Industrial Research Chair program.}
}
\maketitle

\begin{abstract}
In this paper, we propose a spectral-efficient LoRa (SE-LoRa) modulation scheme with a low complexity successive interference cancellation (SIC)-based detector. The proposed communication scheme significantly improves the spectral efficiency of LoRa modulation, while achieving an acceptable error performance compared to conventional LoRa modulation, especially in higher spreading factor (SF) settings. We derive the joint maximum likelihood (ML) detection rule for the SE-LoRa transmission scheme that turns out to be of high computational complexity. To overcome this issue, and by exploiting the frequency-domain characteristics of the dechirped SE-LoRa signal, we propose a low complexity SIC-based detector with a computation complexity at the order of conventional LoRa detection. By computer simulations, we show that the proposed SE-LoRa with low complexity SIC-based detector can improve the spectral efficiency of LoRa modulation up to $445.45\%$, $1011.11\%$, and $1071.88\%$ for SF values of $7$, $9$, and $11$, respectively, while maintaining the error performance within less than $3$ dB of conventional LoRa at symbol error rate (SER) of $10^{-3}$ in Rician channel conditions.
\end{abstract}
\begin{IEEEkeywords}
LoRa, Spectral Efficiency, SIC
\end{IEEEkeywords}

\section{Introduction}
\label{intro}
\IEEEPARstart{T}{he} internet of things (IoT) represents the connection and communication between IoT devices, also known as end-devices (EDs), with the goal of gathering data, processing it, and using the insights obtained to improve daily living through more intelligent and automated decision-making. Developing effective IoT systems is made more challenging by the quick expansion of IoT devices and the growing demand for connectivity. It is still difficult to create adaptable, economical, and energy-efficient systems that can control data transfer, facilitate non-time-sensitive communication, and preserve dependability in the face of scale constraints. Low-power wide area network (LPWAN) technologies have emerged as the primary connectivity choice for IoT networks in order to address these issues. LPWAN is ideal for applications that can accept latency and have restricted data throughput, including battery-powered, low-energy devices, because it provides a wide communication range, energy efficiency, and low implementation costs \cite{Stusek_2022}.

Among the several LPWAN technologies, Semtech's long-range wide area network (LoRaWAN) protocol, which is based on LoRa modulation (also known as chirp spread spectrum (CSS) modulation), has been quickly gaining industrial acceptance throughout the world. LoRaWAN is a popular option for a variety of LPWAN applications since it is an open-source technology \cite{Vangel_2015} that enables independent and inexpensive network implementation. LoRaWAN has been widely used in many different systems and applications. LoRaWAN's IoT EDs function at comparatively low data rates, usually on the order of a few kilobits per second, despite its ability to attain communication ranges of several tens of kilometers and device battery lifetimes of up to ten years \cite{Milarokostas_2023}. For example, a spreading factor (SF) of $7$ with a $500$ kHz bandwidth yields a maximum attainable data rate of $21.9$ kbit/s \cite{9} in the U.S. region. Higher data rate use cases like emergency response, multimedia IoT systems, video streaming, and disaster monitoring are severely limited by this data rate, even though it would be adequate for current LoRaWAN applications. Therefore, improving LoRa modulation's spectral efficiency (SE) is crucial to satisfying present and future IoT demands that call for greater data throughput.

Several studies in the literature have attempted to improve the spectral efficiency of LoRa modulation either by pre-processing the LoRa signal waveform \cite{Hanif_2021_SSK,Hanif_2021_IM,Baruffa_2021,Ma_2021_FBI,Index_ICS} or by adding additional hardware such as multiple antennas at both the transmitter and the receiver to create a multiple-input multiple-output (MIMO) configuration to increase the data rate \cite{Kang_2022,Kang_MIMO_2023,Maleki_MIMO}. However, both approaches introduce a trade-off between improving the performance of LoRa modulation and maintaining its low-cost implementation \cite{Index_ICS}. Recently, a superposition-based LoRa transmission scheme was proposed in \cite{superlora}, based on which, a sequence of LoRa chirps in a LoRa frame can be transmitted in a shorter duration compared to the conventional LoRa transmission to enhance data throughput. However, on the receiver side, the error performance is significantly degraded due to additional interference from preceding and succeeding LoRa signals, resulting in an error floor in all simulation and experimental scenarios.

In this paper, inspired by the work in \cite{superlora}, we propose a spectral-efficient LoRa transmission scheme called SE-LoRa, in which consecutive LoRa chirps within a LoRa frame are transmitted every $\tau<T$ sec ($T$ is the conventional LoRa chirp transmission time). However, by doing so, the received signal within the duration of $T$ will contain a superposition of multiple LoRa chirps: one complete LoRa chirp corresponding to the desired chirp in that specific receiving window and the other truncated LoRa chirps acting as interference. To address this issue, we propose a low complexity detector, which operates based on successive interference cancellation (SIC) and effectively removes the interference caused by preceding and succeeding LoRa signals. By means of simulation, we also show that the proposed SIC-based detector achieves an acceptable error performance. The main contributions of this paper are summarized as follows:
\begin{enumerate}
    \item Derivation of a joint maximum likelihood (ML) detector for SE-LoRa transmission: By presenting SE-LoRa, we first derive the optimal joint ML detector to detect the LoRa chirps within a LoRa frame, jointly. 
    \item Design of a low complexity SIC-based detector: Since the joint ML detector exhibits high computational complexity, especially for practical SF values, we propose a low complexity SIC-based detector based on the frequency-domain characteristics of the dechirped SE-LoRa signal that effectively mitigates inter-chirp interference. 
    \item Simulation-based performance evaluation: The simulation results confirm that the proposed SE-LoRa system, when combined with the low complexity SIC-based detector, maintains desirable error performance while achieving a substantial gain in spectral efficiency compared to conventional LoRa. For instance, our proposed low complexity SIC-based detector enhances the spectral efficiency of LoRa modulation up to $445.45\%$, $1011.11\%$, and $1071.88\%$ for SF values of $7$, $9$, and $11$, respectively, while keeping the error performance within less than $3$ dB of conventional LoRa at symbol error rate (SER) of $10^{-3}$ in Rician channel conditions.
\end{enumerate}

The rest of the paper is organized as follows. Section II presents the state-of-the-art by discussing the related works. In Section III, the system model is introduced along with the SE-LoRa transmission scheme. Section IV derives the joint ML detection rule and discusses its computational complexity. The low complexity SIC-based detector is presented in Section~V. Simulation results are provided in Section VI. Finally, in Section VIII, the paper is concluded. 
\section{Related Works}
\label{relate}
Slope-shift keying LoRa (SSK-LoRa) is a modulation introduced in the work of \cite{Hanif_2021_SSK}. This method extends the conventional LoRa symbol set by using both up-chirps, whose frequency increases linearly with time, and down-chirps, whose frequency decreases linearly with time. One extra bit can be transmitted during each symbol period thanks to this extension. Although the spectral efficiency improvement of SSK-LoRa is modest, e.g., $14.3\%$ for ${\rm SF}=7$, the performance gain in SSK-LoRa is around $0.35$ dB compared to the conventional LoRa method at a bit error rate (BER) of $10^{-4}$.

Index modulation (IM) is another approach that has drawn interest as a successful technique for increasing data rates in different modulation schemes \cite{Ishikawa_2018}. Frequency-shift chirp spread spectrum with index modulation (FSCSS-IM) is a suggested combination of IM and LoRa modulation in \cite{Hanif_2021_IM}. Each transmitted signal in this scheme is generated by adding several unique LoRa waveforms, each of which represents a portion of the communicated data. Consequently, each symbol transmits a higher total number of bits, e.g., equivalent to $157\%$ spectral efficiency improvement compared to conventional LoRa for ${\rm SF}=7$. The ideal detector for this transmission scheme needs to scan through every possible combination of LoRa symbols utilized at the transmitter. In \cite{Hanif_2021_IM}, a lower-complexity receiver structure is also added to lower this complexity. The simulation results demonstrate that while FSCSS-IM produces a noticeable increase in spectral efficiency, its BER performance is somewhat worse than that of traditional LoRa modulation, e.g., near $0.3$ dB degradation at BER of $10^{-4}$ for ${\rm SF}=8$ in an additive white Gaussian noise (AWGN) channel condition. 

To further enhance the spectrum efficiency of LoRa systems, the authors in \cite{Baruffa_2021} present a modulation approach called in-phase and quadrature chirp index modulation (IQCIM), building on the integration of FSCSS-IM and IQ multiplexing. Two FSCSS-IM waveforms, which function as the in-phase and quadrature components, are combined to create the transmitted signal in this approach. Thus, with the same bandwidth and system characteristics, IQCIM achieves twice the bit rate of FSCSS-IM. According to performance evaluation results, IQCIM maintains equivalent computational complexity while dramatically improving spectral efficiency when compared to traditional LoRa modulation. Only a small increase in energy consumption of roughly $0.9$ dB is required to achieve this improvement. 

In \cite{Ma_2021}, the authors examine the use of multiple antennas to improve the link dependability of LoRa-based communication when MIMO setups are combined with LoRa modulation. Two short studies \cite{Kang_2022,Kang_MIMO_2023}, which describe precoding techniques and detection frameworks for MIMO-LoRa systems targeted at high-data-rate IoT applications, provide more developments. To reduce interference between the broadcast signals, each transmit antenna in these methods transmits a LoRa waveform with a unique SF. This configuration benefits from spatial diversity at the receiver side while increasing the spectral efficiency of LoRa transmission in proportion to the number of transmit antennas. According to simulation results in \cite{Kang_2022,Kang_MIMO_2023}, the suggested MIMO-LoRa system performs much better than conventional LoRa in terms of feasible data rate and BER. As an example, with a $4\times4$ MIMO configuration, MIMO-LoRa improves the performance of LoRa modulation by $6$ dB in AWGN channel condition at BER of $10^{-4}$, while also improving the spectral efficiency by $240\%$. However, because ML detection needs to be done for every conceivable combination of symbols, the detection process is still computationally demanding. As the most recent work in MIMO-LoRa configurations, the authors in \cite{Maleki_MIMO} introduce MIMO-CSS-PMM, a MIMO configuration for LoRa modulation in conjunction with permutation matrix modulation (PMM). The suggested method improves LoRa transmission's error performance as well as SE. For instance, for the $4\times4$ MIMO configuration, MIMO-CSS-PMM scheme improves the spectral efficiency of LoRa modulation by $340\%$, while improving the performance by $9$ dB in a Rayleigh fading channel condition, at BER of $10^{-4}$. Although an ideal ML detector is derived, it turns out to be computationally demanding. Two semi-coherent detection approaches are suggested in order to simplify the detection procedure. According to simulation results, the proposed detectors greatly reduce complexity while achieving BER performance close to conventional LoRa modulation. However, the performance gains achieved through MIMO configurations come at the expense of additional hardware complexity and cost.

Most recently, the work of \cite{Index_ICS} presents a new modulation approach that uses indexing interleaved chirp spreading (ICS) LoRa symbols to transmit information. The suggested technique offers a greater data rate while maintaining the constant-envelope characteristic of conventional LoRa modulation. For the index-interleaved framework, a low-complexity detection algorithm is created, and the associated BER expression is obtained analytically. The suggested Index-ICS LoRa scheme outperforms traditional LoRa and ICS-LoRa by roughly $28.6\%$ and $14.3\%$, respectively, for an SF value of $7$. The approach achieves better throughput that comes at a cost of error performance degradation compared to conventional LoRa modulation, e.g., $2.5$ dB at the BER of $10^{-3}$ for a Rayleigh fading channel, according to simulation data.

\section{System Model}
\label{sysmod}
In the following, we discuss the basics of LoRa modulation and its conventional detection process. Afterwards, we introduce the SE-LoRa transmission scheme.
\subsection{Basics of LoRa Modulation and Conventional Detection}
Consider a LoRa-based communication system architecture in which a LoRa ED transmits to a stationary terrestrial LoRaWAN GW. The LoRa ED is set to transmit with the LoRaWAN specification as the spreading factor of $\rm SF\in\{7,8,9,10,11,12\}$ and bandwidth of $\rm B\in\{125,250,500\}$ kHz \cite{Maleki_chirp}. The LoRa modulation is modeled as a set of $M = 2^{\rm SF}$ orthogonal chirps, each of which modulates a binary sequence with a length equal to $\rm SF$. The system involves mapping each information symbol onto a chirp waveform whose frequency increases linearly over the bandwidth. The key property of this chirp is that, once the symbol frequency reaches its maximum, it wraps around to the minimum frequency and continues sweeping until the entire bandwidth has been covered once \cite{Vangelista_2017,Maleki_chirp}. 

Assume that we have a binary sequence of $b_0b_1\dots b_{{\rm SF}-1}$. With LoRa modulation, first, the decimal value of $s=\sum^{{\rm SF}-1}_{i=0}b_i2^i$ is obtained. Obviously, we have $s\in\mathcal{S}=\{0,1,\dots,M-1\}$. Then, by dividing the bandwidth into $M$ frequency bins, the frequency of $-{\rm B}/2+{\rm B}s/M$ is selected as the chirp start frequency, i.e., the information-bearing element. The symbol duration of LoRa modulation can be calculated as $T=M/{\rm B}$ \cite{Maleki_chirp}. Hence, the spectral efficiency of the LoRa modulation can be formulated as:
\begin{IEEEeqnarray}{rCl}
\label{SELoRa}
\eta_{\rm L}=\frac{R_{\rm L}}{\rm B} = \frac{{\rm SF}/T}{{\rm B}} = \frac{{\rm SF}}{M}=\frac{{\rm SF}}{2^{\rm SF}}, 
\end{IEEEeqnarray}
where $R_{\rm L}={\rm SF}/T$ is the LoRa modulation bit rate. Referring to equation (\ref{SELoRa}), it is evident that as $\rm SF$ increases, the denominator of $\eta_{\rm L}$ increases at a faster rate than the numerator, leading to a reduction in SE. 

The baseband representation of a continuous-phase discrete-time LoRa signal for symbol $s$ can be expressed as \cite{Maleki_chirp}:
\begin{IEEEeqnarray}{rCl}
\label{xdisc}
x_{\rm L}[n;s]&=&\exp\left(2\pi j\frac{n^2+2ns-nM}{2M}\right)g_{M}[n]\nonumber\\
&=&x_{\rm L}[n;0]\exp\left(2\pi j\frac{ns}{M}\right)g_{M}[n],
\end{IEEEeqnarray}
where $g_{M}[n]=1$ for $n=0,1,\dots,M-1$ and zero otherwise. In the literature, $x_{\rm L}[n;0]=\exp(2\pi j (n^2-nM)/(2M))$ is referred to as the basic up-chirp \cite{Vangelista_2017,Maleki_chirp} modulating the LoRa symbol $s=0$, and is used as part of the LoRa demodulation procedure.

LoRa modulation has a unique frequency-domain characteristic by which the demodulation can be performed using dechirping (multiplying the received signal by the complex conjugate of the basic up-chirp, i.e., $x_{\rm L}^*[n;0]$ \cite{Maleki_chirp}) and discrete Fourier transform (DFT) operations. As shown in \cite{Maleki_chirp}, this detection method stems directly from the ML detection of a LoRa chirp. The conventional coherent LoRa detection for a received signal of $r[n]=\sqrt{P}hx_{\rm L}[n;s]+w[n]$ ($n=0,1,\dots,M-1$) can be formulated as:
\begin{IEEEeqnarray}{rCl}
\label{LoRadet}
\hat{s}=\argmax_{u\in\mathcal{S}}\mathcal{R}\left\{h^*\times \underbrace{{\rm DFT}\left\{r[n]x_{\rm L}^*[n;0]\right\}}_{Y[u]}\right\},
\end{IEEEeqnarray}
where $P$, $h$, and $w[n]$ denote the LoRa ED transmitting power, the complex channel gain, and AWGN. Also, $\mathcal{R}\{x\}$ is the real part of complex value $x$. It should be noted that in this paper, we assume that the channel state information (CSI) is known at the receiver. One way for the receiver to obtain this information is by estimating the channel coefficient using the method proposed in \cite{Almeida_2020}. In this approach, the receiver estimates the channel coefficients exploiting the preamble part of the LoRa frame, which is transmitted at the start of each frame for synchronization purposes. The LoRa symbols in the preamble are known at the receiver side. Hence, they can be exploited as pilot signals for channel estimation. This technique can estimate the channel coefficients with a very high accuracy and without any additional signaling overhead \cite{Almeida_2020,Maleki_MIMO}. Moreover, the technique proposed in \cite{khai_2021} can also be used to estimate the channel coefficients using the previously detected LoRa symbols. Note that after estimating the channel coefficients, the channel is assumed to remain constant throughout the transmission of a LoRa frame. This assumption is reasonable for typical LoRa applications, where EDs and gateways (GWs) are largely stationary.

\subsection{SE-LoRa Transmission}
\begin{figure}[t]
  \centering
\includegraphics[width=0.9\linewidth]{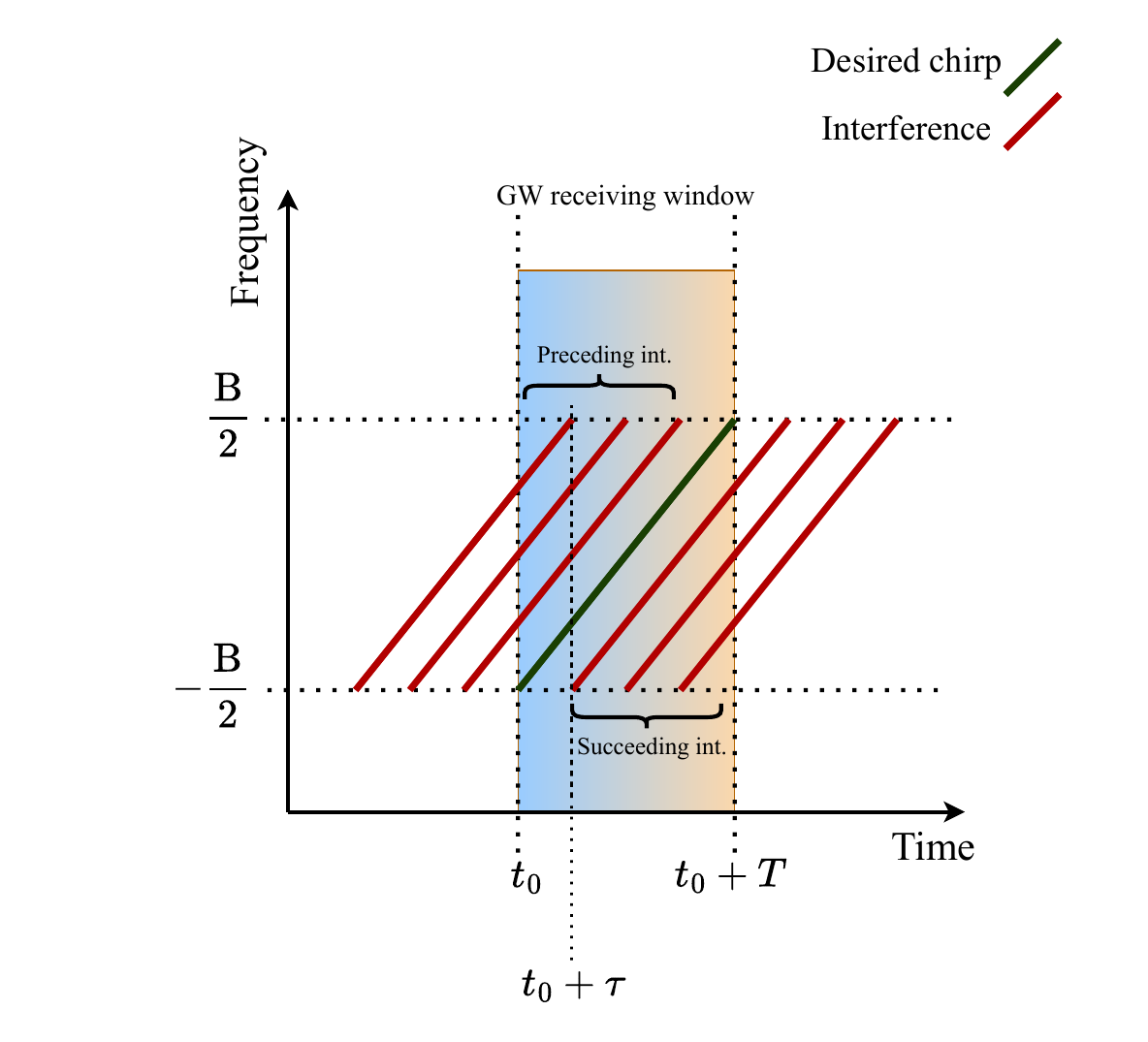}
  \caption{SE-LoRa transmission scheme for $\tau=T/4$ (Note that for illustration purposes, we set $s=0$ for all of the presented chirps)}
  \label{SLchirp}
\end{figure}

The SE-LoRa transmission scheme is presented in this paper, inspired by the work of \cite{superlora} to improve the spectral efficiency in LoRaWAN-based IoT networks. The basis of this modulation is to transmit LoRa signals every $\tau =T/K$ seconds, where $K\geq1\in \mathbb{Z}$ ($\mathbb{Z}$ is the set of integers). Thus, at the receiver, the GW observes a superposition of the desired LoRa signal and several truncated LoRa signals acting as interference, within a time interval of $T$. For illustration, Fig. \ref{SLchirp} depicts this scenario with $\tau$ set to $T/4$. In this case, within a reception window of duration $T$, the GW captures the complete LoRa signal, referred to in this paper as the desired chirp, along with interference originating from $3$ preceding and $3$ succeeding truncated LoRa signals. It also should be noted that exploiting this approach, the receiver needs to adjust its receiving windows to shift by $\tau$ instead of $T$ to sequentially detect the consecutive LoRa symbols within a LoRa frame. For example, the reception window for the first LoRa symbol spans from $0$ to $T$, for the second symbol from $\tau$ to $T+\tau$, and so on.

To formulate the spectral efficiency improvement provided by SE-LoRa modulation, it can be seen that for transmission of $l$ LoRa symbols using conventional LoRa modulation, the ED requires a total time of $lT$. However, exploiting SE-LoRa, the time required for transmission of the same amount of information is $(l-1)\tau+T$. Hence, the spectral efficiency of the SE-LoRa modulation can be represented as
\begin{IEEEeqnarray}{rCl}
\label{SESELoRa}
\eta_{\rm SE-L}=\frac{R_{\rm SE-L}}{\rm B} = \frac{Kl}{K+l-1}\eta_L. 
\end{IEEEeqnarray}
It can be seen that for typical LoRa frame payloads, where $l$ is relatively large ($l\gg K$), we have $\eta_{\rm SE-L}\approx K\eta_{\rm L}$.

We can formulate the sampled received SE-LoRa signal within one receiving window with the duration of $T$ (e.g. the first receiving window $0\leq t=nT/M< T$ without the loss of generality) as:
\begin{IEEEeqnarray}{rCl}
\label{r}
r[n]=\sqrt{P}hX_{\rm SE}[n;\left\{s_i\right\}_{i=-K+1}^{K-1}]+w[n],
\end{IEEEeqnarray}
for $n=0,1,\dots,M-1$. Moreover, $X_{\rm SE}[n]$ represents the superposition of LoRa signals (desired chirp modulating LoRa symbol $s_0$, $K-1$ preceding truncated chirps modulating $\{s_i\}_{i=-K+1}^{-1}$, and $K-1$ succeeding truncated chirps modulating $\{s_i\}_{i=1}^{K-1}$) received within the GW receiving window and can be expressed as:
\begin{IEEEeqnarray}{rCl}
\label{SEX}
X_{\rm SE}[n;\left\{s_i\right\}_{i=-K+1}^{K-1}]&=&\underbrace{x_{\rm L}[n;s_0]}_{\rm Desired\;chirp} \nonumber\\
&&+ \underbrace{\sum_{i=-K+1}^{-1}x_{\rm L}[n-i\lambda;s_{i}]g_{M}[n]}_{\rm Preceding\;interference}\nonumber\\
&&+\underbrace{\sum_{i=1}^{K-1}x_{\rm L}[n-i\lambda;s_{i}]g_{M}[n]}_{\rm Succeeding\;interference},
\end{IEEEeqnarray}
where $\lambda=\lfloor M/K \rfloor$ ($\lfloor . \rfloor$ is the floor operator). Also, for the ease of future derivations, we expand $x_{\rm L}[n-i\lambda;s_i]$ using the basic up-chirp expression as follows:
\begin{IEEEeqnarray}{rCl}
\label{xLrew}
x_{\rm L}[n-i\lambda;s_i] &=& x_{\rm L}[n;0]\exp\Bigg{[}2\pi j\frac{\alpha(i,\lambda,s_i)n+\beta(i,s_i,\lambda)}{2M}\Bigg{]}\nonumber\\
&&\times g_{M}[n-i\lambda],
\end{IEEEeqnarray}
in which we have: 
\begin{IEEEeqnarray}{rCl}
\label{alphabeta}
\begin{cases}
\alpha(i,\lambda,s_i)= 2s_i-2i\lambda,\\
               \beta(i,s_i,\lambda) = i^2\lambda^2-2i\lambda s_i+i\lambda M.
              \end{cases}
              \end{IEEEeqnarray}

It should be noted that the SE-LoRa scheme cannot readily be deployed on typical commercial-off-the-shelf (COTS) LoRa transmitters, because it depends on accessing the modulator’s in-phase/quadrature (I/Q) samples. Conceptually, the transmitter is no longer a standard LoRa modem that transmits one conventional LoRa chirp waveform per $T$; instead, it must generate an SE-LoRa signal in which multiple time-shifted LoRa signal samples are superimposed and then transmitted over one symbol interval of $T$, as can be inferred from (\ref{SEX}). Since common LoRa chipsets implement modulation/demodulation internally and do not expose the necessary baseband interfaces, achieving this superposition generally cannot be accomplished through a simple firmware update.

In the following section, we derive the joint ML detection rule for SE-LoRa modulation, by which for a LoRa frame, one can detect the sequence of LoRa symbols forming the payload.
\section{Joint ML Detection}
\begin{figure*}[t]
  \centering
\includegraphics[width=\linewidth]{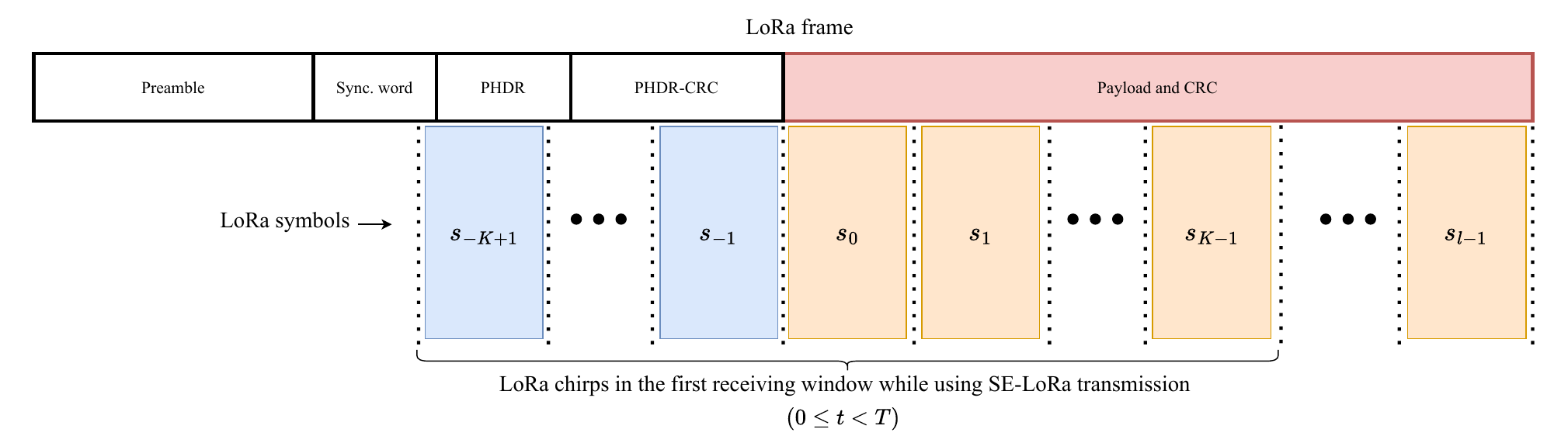}
  \caption{LoRa frame structure}
  \label{SELoRaFrame}
\end{figure*}

The term ``joint'' here refers to the detection of all LoRa symbols within a LoRa frame payload. A LoRa frame is a consecutive sequence of LoRa chirps, modulating LoRa symbols. As illustrated in Fig. \ref{SELoRaFrame}, it can be seen that the LoRa payload is preceded by multiple parts, i.e., physical header (PHDR), PHDR cyclic redundancy check (CRC), synchronization word, and preamble. It should be noted that these parts are known at the receiver, i.e., $s_{-1}$, $\dots$, and $s_{-K+1}$ are known. Moreover, at the receiver side, we set $t=0$ as the time that the receiver starts receiving the payload of the frame after the synchronization process. 

Now, consider transmission of an entire LoRa frame with the SE-LoRa scheme. Hence, we can present the received SE-LoRa payload ($s_0$, $s_1$, $\dots$, $s_{l-1}$) samples for the time duration of $0\leq t\leq lT$ as:
\begin{IEEEeqnarray}{rCl}
\label{RxFrame}
r_{\rm F}[n]&=&\underbrace{\sqrt{P}h\sum_{i=-K+1}^{-1}x_{\rm L}[n-i_1\lambda;s_{i}]g_{M}[n]}_{C}\nonumber\\
&&+\sqrt{P}h\underbrace{\sum_{i=0}^{l-1}x_{\rm L}[n-i\lambda;s_{i}]}_{X_{\rm SE-F}[n;\left\{s_i\right\}_{i=0}^{l-1}]}+w[n],
\end{IEEEeqnarray}
for $n=0,1,\dots,lM-1$. As mentioned, the term $C$, is known at the receiver \cite{9}. Using the Bayes rule \cite{Gold_2005}, we can write: 
\begin{IEEEeqnarray}{rCl}
\label{bayes}
p\left(\left\{s_i\right\}_{i=0}^{l-1}|\mathbf{r}_{\rm F},h\right) &=&p\left(\mathbf{r}_{\rm F}|\left\{s_i\right\}_{i=0}^{l-1},h\right)\frac{p\left(\left\{s_i\right\}_{i=0}^{l-1}\right)}{p\left(\mathbf{r}_{\rm F}|h\right)},\nonumber\\
\end{IEEEeqnarray}
where $\mathbf{r}_{\rm F}=[r_{\rm F}[0],r_{\rm F}[1],\dots,r_{\rm F}[lM-1]]^\top$. Assuming perfect CSI at the receiver, and equally likely LoRa symbols, we can obtain the joint ML detection rule as:
\begin{IEEEeqnarray}{rCl}
\label{ML1}
\left\{\left\{\hat{s}_i\right\}_{i=0}^{l-1}\right\}=\argmax_{\left\{{s}_i\right\}_{i=0}^{l-1} \in \mathcal{S}}p\left(\mathbf{r}_{\rm F}|\left\{s_i\right\}_{i=0}^{l-1},h\right).\nonumber\\
\end{IEEEeqnarray}
Considering that we have $w[n]\sim \mathcal{CN}(0,\sigma^2)$, the objective function of the joint ML detection rule in (\ref{ML1}) can be written as:
\begin{IEEEeqnarray}{rCl}
\label{obj}
p\Big{(}\mathbf{r}_{\rm F}&|&\{s_i\}_{i=0}^{l-1},h\Big{)}=\frac{1}{\pi^{lM}\sigma^{2lM}} \nonumber\\
&&\times \exp\left\{-\sum_{n=0}^{lM-1}\frac{\big{|}r_{\rm F}[n]-h\sqrt{P}X_{\rm SE-F}[n;\left\{s_i\right\}_{i=0}^{l-1}]\big{|}^2}{\sigma^2}\right\}.\nonumber\\
\end{IEEEeqnarray}
After some mathematical simplifications, we obtain:
\begin{IEEEeqnarray}{rCl}
\label{exp}
p\Big{(}\mathbf{r}_{\rm F}&&|\left\{s_i\right\}_{i=0}^{l-1},h\Big{)}=\zeta\nonumber\\
&&\times\exp\Bigg{(}2\sqrt{P}\sum_{n=0}^{lM-1}\mathcal{R}\left\{h^*r_{\rm F}[n]X_{\rm SE-F}^*[n;\left\{s_i\right\}_{i=0}^{l-1}]\right\}\nonumber\\
&&-P|h|^2\sum_{n=0}^{lM-1}X_{\rm SE-F}[n;\left\{s_i\right\}_{i=0}^{l-1}]X_{\rm SE-F}^*[n;\left\{s_i\right\}_{i=0}^{l-1}]\Bigg{)},\nonumber\\
\end{IEEEeqnarray}
where $\zeta$ is the constant term expressed as:
\begin{IEEEeqnarray}{rCl}
\label{zeta}
\zeta=\frac{1}{\pi^{lM}\sigma^{2lM}}\exp\left(-\sum^{lM-1}_{n=0}|r_{\rm F}[n]|^2\right).
\end{IEEEeqnarray}
Accordingly, by substituting (\ref{exp}) into (\ref{ML1}), taking the logarithm, and dropping the constant term, the final joint ML detection rule can be expressed as:
\begin{IEEEeqnarray}{rCl}
\label{MLf}
\Big{\{}\{&\hat{s}&_i\}_{i=0}^{l-1}\Big{\}}=\argmax_{\left\{{s}_i\right\}_{i=0}^{l-1} \in \mathcal{S}}2\sqrt{P}\nonumber\\
&&\times\sum_{n=0}^{lM-1}\mathcal{R}\left\{h^*r_{\rm F}[n]X_{\rm SE-F}^*[n;\left\{s_i\right\}_{i=0}^{l-1}]\right\}\nonumber\\
&&-P|h|^2\sum_{n=0}^{lM-1}X_{\rm SE-F}[n;\left\{s_i\right\}_{i=0}^{l-1}]X_{\rm SE-F}^*[n;\left\{s_i\right\}_{i=0}^{l-1}].\nonumber\\
\end{IEEEeqnarray}
\subsection{Joint ML Detection Computational Complexity}
From (\ref{MLf}), it can be seen that exhaustive search can be performed over all possible values of $\{s_i\}_{i=0}^{l-1}$ with a total number of $M^l$ searches. The total number of operations (multiplication and summation) in each search can be calculated as $23lM+3$, considering that all of the parameters are complex values. Therefore, the total complexity of joint ML detection can be expressed as $\mathcal{O}(lM^{l+1})$. It can be seen that for the practical values of $M$, i.e., $\{128,256,512,1024,2048,4096\}$, and even short LoRa payloads in real world situations, e.g., $l=10$ \cite{Hoeller_2018}, the joint ML detection is not practically implementable due to a very high computational burden. Moreover, for the joint ML detection to be carried out, one has to wait for the entire frame to be received, which introduces an unacceptable delay to the detection procedure. Hence, in the following section, we present our low complexity SIC-based detector, which offers significantly lower complexity, making the implementation of SE-LoRa practical in terms of both computational efficiency and error performance.

\section{Proposed low complexity SIC-based SE-LoRa detector}
To gain a better understanding of the foundation of our proposed low complexity SIC-based detector, we first examine the frequency-domain characteristics of the dechirped SE-LoRa signal in the following subsection.
\subsection{Frequency-domain Characteristics of Dechirped SE-LoRa}
\begin{figure}[t]
  \centering
\includegraphics[width=0.8\linewidth]{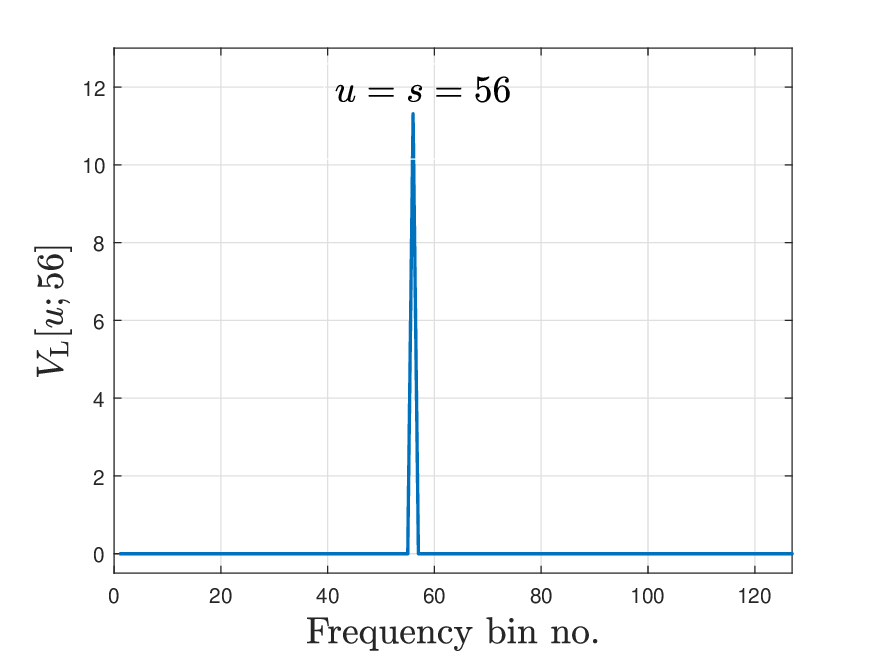}
  \caption{DFT of the dechirped LoRa signal with ${\rm SF}=7$ and $s=56$.}
  \label{LoRapeak}
\end{figure}

To investigate the frequency-domain characteristics of a dechirped LoRa signal, we neglect the effects of AWGN and fading for now and consider a LoRa signal $x_{\rm L}[n;s]$, for which the dechirping and DFT operations are performed as \cite{Maleki_chirp}: 
\begin{IEEEeqnarray}{rCl}
\label{dech}
V_{\rm L}[u;s]&=&{\rm DFT}\left\{\underbrace{x_{\rm L}[n;s]x_{\rm L}^*[n;0]}_{x_{\rm L}^{\rm (dech.)}[n;s]}\right\}\nonumber\\
&=&{\frac{1}{\sqrt M}}\sum_{n=0}^{M-1}\underbrace{\exp{\left\{j2\pi\frac{n\left(s+M-u\right)}{M}\right\}}}_{\text{is periodic with the period of $M$}}\nonumber\\
&=&  \begin{cases}
\sqrt M, & {\rm{if}}\ u=s, \\
                0, & {\rm{otherwise.}}
              \end{cases}
\end{IEEEeqnarray}
This means that the output of the dechirping and DFT operations of a LoRa chirp can be represented as a frequency-domain signal $V[u]$ for $u=0,1,\dots,M-1$ which has a peak at the location (frequency bin) of $u=s$, and is zero otherwise. This is illustrated in Fig. \ref{LoRapeak} for an arbitrary $s=56$ and ${\rm SF}=7$ in which $V_{\rm L}[u;56]$ is zero at all frequency bins except for the frequency bin $56$ with a peak at $\sqrt{2^7}=11.31$.

It should be noted that locating this peak forms the basis for detecting a complete LoRa chirp. However, in the case of SE-LoRa modulation, where truncated versions of LoRa chirps are also involved, the output of the DFT and dechirping operations must be examined. To formulate this, we write the output of the DFT and dechirping operations on $X_{\rm SE}[n;\left\{s_i\right\}_{i=-K+1}^{K-1}]$ (which is defined for $0\leq t=nT/M< T$) as (\ref{dechSE}) at the top of the next page. Note that since we are performing coherent detection with the help of CSI available at the receiver, we are interested in the real part of $V_{\rm SE}[u;\left\{s_i\right\}_{i=-K+1}^{K-1}]$. 

\begin{figure*}
\normalsize
\begin{IEEEeqnarray}{rCl}
\label{dechSE}
V_{\rm SE}[u;\left\{s_i\right\}_{i=-K+1}^{K-1}]&=&{\rm DFT}\left\{\underbrace{X_{\rm SE}[n;\left\{s_i\right\}_{i=-K+1}^{K-1}]b_{0}^*[n]}_{X_{\rm SE}^{\rm (dech.)}[n;\left\{s_i\right\}_{i=-K+1}^{K-1}]}\right\}=\frac{1}{\sqrt M}\sum_{n=0}^{M-1}{X_{\rm SE}^{\rm (dech.)}[n;\left\{s_i\right\}_{i=-K+1}^{K-1}]\exp{\left(-j2\pi \frac{un}{M}\right)}}\nonumber\\
&=& V_{\rm L}[u;s_0]+\frac{1}{\sqrt{M}}\sum_{i=-K+1}^{-1}\sum_{n=0}^{M+\lambda i-1}\exp\left\{j2\pi\left[\frac{\left[\alpha(i,K,s_{i})-2u\right]n+\beta(i,K,s_{i})}{2M}\right]\right\}\nonumber\\
&&+\frac{1}{\sqrt{M}}\sum_{i=1}^{K-1}\sum_{n=\lambda i}^{M-1}\exp\left\{j2\pi\left[\frac{\left[\alpha(i,K,s_{i})-2u\right]n+\beta(i,K,s_{i})}{2M}\right]\right\}.
\end{IEEEeqnarray}
\hrulefill
\vspace
*
{4pt}
\end{figure*}

Consider the following terms in (\ref{dechSE})
\begin{IEEEeqnarray}{rCl}
\label{term1}
\sum_{n=0}^{M+\lambda i-1}\exp\left\{j2\pi\left[\frac{\left[\alpha(i,K,s_{i})-2u\right]n+\beta(i,K,s_{i})}{2M}\right]\right\},\nonumber\\
\end{IEEEeqnarray}
which is the DFT of the dechirped version of the $i$ preceding LoRa chirp, and
\begin{IEEEeqnarray}{rCl}
\label{term2}
\sum_{n=\lambda i}^{M-1}\exp\left\{j2\pi\left[\frac{\left[\alpha(i,K,s_{i})-2u\right]n+\beta(i,K,s_{i})}{2M}\right]\right\},\nonumber\\
\end{IEEEeqnarray}
which is the DFT of the dechirped version of the $i$ succeeding LoRa chirp.

By taking the derivative of the exponential expressions in (\ref{term1}) and (\ref{term2}) and equating it to zero, it can be seen that the locations of the peaks corresponding to these terms can be obtained by setting $\alpha(i,K,s_{i})-2u=0$, i.e., $2s_i-2i\lambda-2u=0$. Hence, it can be concluded that the peak locations depend on the values of $s_i$, $i$, and $\lambda$. Also, at the peak locations, the terms in (\ref{term1}) and (\ref{term2}) are reduced to
\begin{IEEEeqnarray}{rCl}
\label{term1amp}
\sum_{n=0}^{M+\lambda i-1}\exp\left\{j2\pi\left[\frac{\beta(i,K,s_{i})}{2M}\right]\right\},
\end{IEEEeqnarray}
and
\begin{IEEEeqnarray}{rCl}
\label{term2amp}
\sum_{n=\lambda i}^{M-1}\exp\left\{j2\pi\left[\frac{\beta(i,K,s_{i})}{2M}\right]\right\}.
\end{IEEEeqnarray}
By calculating the absolute value of (\ref{term1amp}) and (\ref{term2amp}), the magnitudes of these functions at the peak locations can be obtained as $M-\lambda|i|$.

Accordingly, there are several insights that can be interpreted from (\ref{dechSE}), listed as follows:
\begin{itemize}
    \item There is a positive peak in $\mathcal{R}\{V_{\rm SE}[u;\left\{s_i\right\}_{i=-K+1}^{K-1}]\}$ at the location of frequency bin $s_0$, corresponding to the desired chirp, with the amplitude of $\sqrt{M}$ (resulted from the term $V_{\rm L}[u;s_0]$ in (\ref{dechSE})).
    \item By setting the term $\left[\alpha(i,K,s_{i})-2u\right]n+\beta(i,K,s_{i})$ in (\ref{dechSE}) equal to zero, we obtain a maximum of $2K-1$ distinct peak locations in $\mathcal{R}\{V_{\rm SE}[u;\left\{s_i\right\}_{i=-K+1}^{K-1}]\}$ (positive or negative depending on the values of $i$, $K$, and $s_{i}$), corresponding to the preceding ($i=-K+1,\dots,-1$) and the succeeding ($i=1,\dots,K-1$)  LoRa chirps.
    \item The peak locations can be expressed using the following equation:
\begin{IEEEeqnarray}{rCl}
\label{locpeak}
\left\{p_{i}\right\}_{i=-K+1}^{K-1}=\mod\left(\left\{s_{i}\right\}_{i=-K+1}^{K-1}-i\lambda,M\right).\nonumber \\
\end{IEEEeqnarray}
    \item The magnitude of the peaks resulting from the desired, preceding, and succeeding LoRa chirps, appearing in $|V_{\rm SE}[u;\left\{s_i\right\}_{i=-K+1}^{K-1}]|$, can be calculated as $(M-\lambda |i|)/\sqrt{M}$. 
\end{itemize}
Note that since $\left\{s_{i}\right\}_{i=-K+1}^{K-1}$ are generated randomly, the $2K-1$ peak locations $\{p_i\}_{i=-K+1}^{K-1}$ might not be unique.

\begin{figure}[t]
\centering
\includegraphics[width=0.8\linewidth]{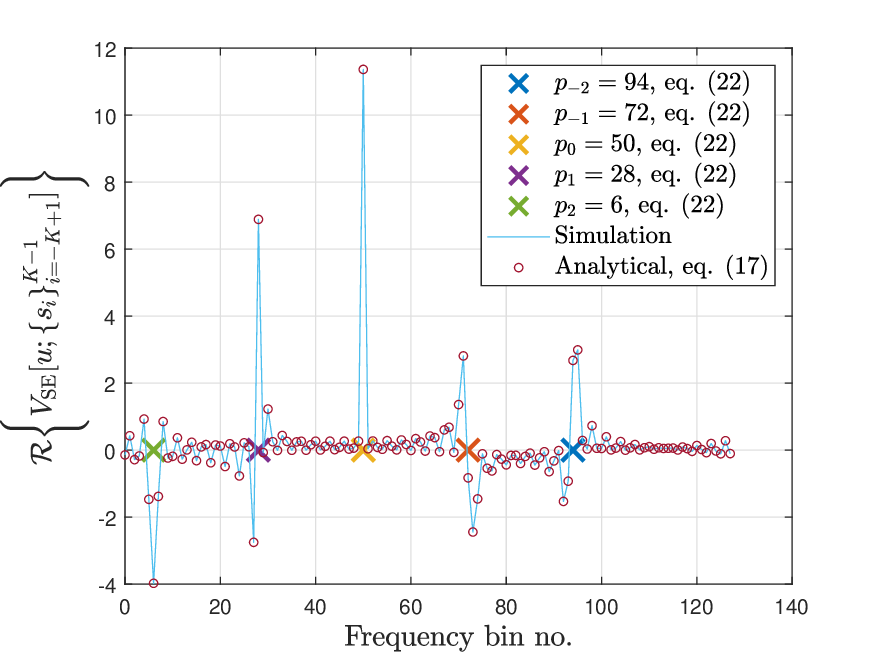}
\caption{Real part of DFT of the dechirped SE-LoRa signal with $K=3$, $\rm SF=7$, and arbitrary LoRa symbols of $s_{-2}=10$, $s_{-1}=30$, $s_{0}=50$, $s_{1}=70$, and $s_{2}=90$.}
  \label{reVSE}
\end{figure}

\begin{figure}[t]
\centering
\includegraphics[width=0.8\linewidth]{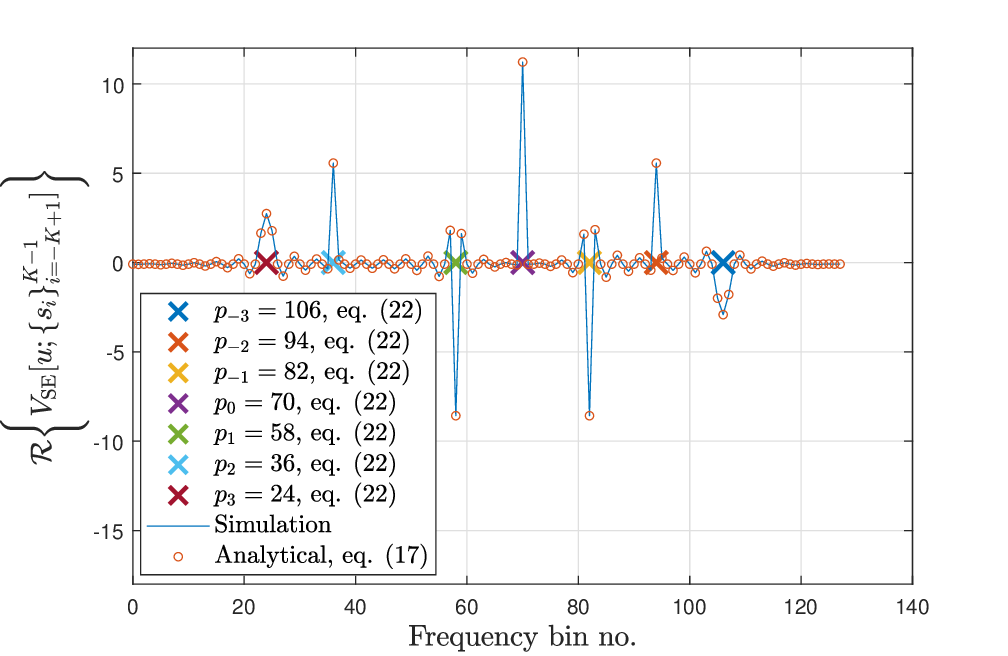}
\caption{Real part of DFT of the dechirped SE-LoRa signal with $K=4$, $\rm SF=7$, and arbitrary LoRa symbols of $s_{-3}=10$, $s_{-2}=30$, $s_{-1}=50$, $s_{0}=70$, $s_{1}=90$, $s_{2}=100$, and $s_{3}=120$.}
  \label{reVSEK4}
\end{figure}

To further illustrate the discussed insights, Fig. \ref{reVSE} and Fig. \ref{reVSEK4} are presented. In Fig. \ref{reVSE}, the real part of $V_{\rm SE}[u;\left\{s_i\right\}_{i=-K+1}^{K-1}]$ is shown for $K=3$ and ${\rm SF}=7$, based on both simulation results and analytical expression in (\ref{dechSE}). The example LoRa symbol sequence is chosen as $s_{-2}=10$, $s_{-1}=30$, $s_{0}=50$, $s_{1}=70$, and $s_{2}=90$. It can be seen that the desired chirp has a positive peak at the location $p_0=50$. Moreover, the peak locations of $p_{-2}=94$, $p_{-1}=72$, $p_{1}=28$, and $p_{2}=6$ contain positive/negative values at or around the peak location. 

Similarly, Fig. \ref{reVSEK4} presents the real part of $V_{\rm SE}[u;\left\{s_i\right\}_{i=-K+1}^{K-1}]$ for $K=4$ and ${\rm SF}=7$, obtained from both simulation results and analytical expression in (\ref{dechSE}). The example LoRa symbol sequence is chosen as $s_{-3}=10$, $s_{-2}=30$, $s_{-1}=50$, $s_{0}=70$, $s_{1}=90$, and $s_{2}=100$, $s_3=120$. The desired chirp has a positive peak at the location $p_0=70$. Moreover, the peak locations of $p_{-3}=106$, $p_{-2}=94$, $p_{-1}=82$, $p_{1}=58$, $p_{2}=36$, and $p_3=24$ contain positive or negative values exactly at the peak location. It should be noted that the validity of equations (\ref{dechSE}) and (\ref{locpeak}) can be confirmed using the results provided in Fig. \ref{reVSE} and Fig. \ref{reVSEK4}. Moreover, from Fig. \ref{reVSE} and Fig. \ref{reVSEK4}, it can be observed that the interfering peaks have sharp positive or negative shapes when $K=4$. However, for the case of $K=3$, these peaks exhibit both negative and positive parts around the peak location. This directly affects the error performance of the proposed SIC-based detector, as will be discussed later in Section~VI.

\subsection{SIC-based Detection}
\begin{figure}[htp]
  \centering
\subfigure[Error scenario 1]
    {
        \includegraphics[width=0.8\linewidth]{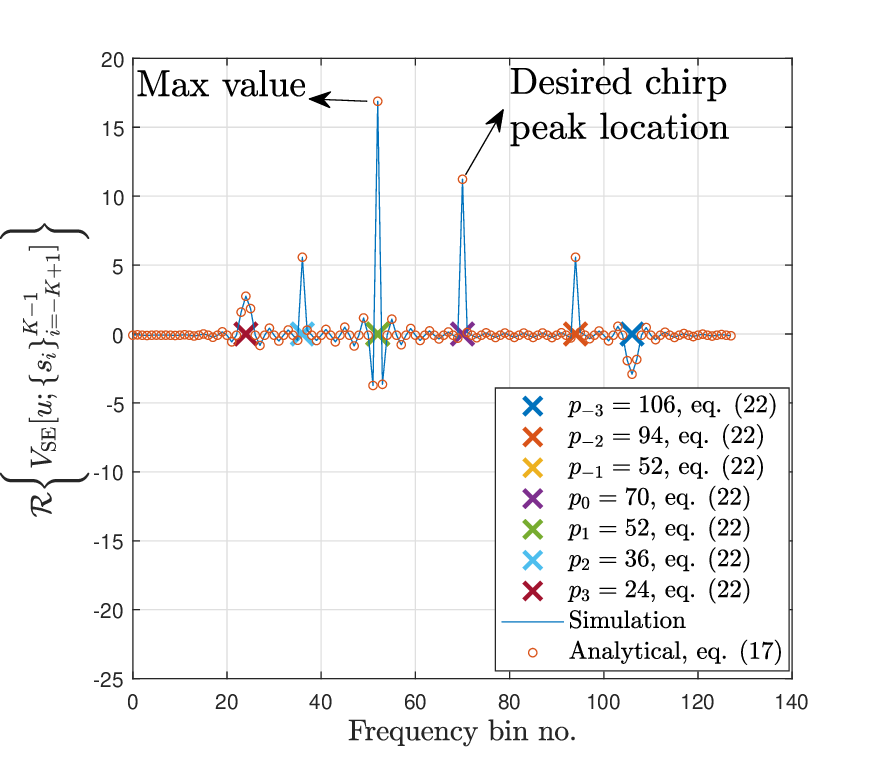}
        
    }
\\

\subfigure[Error scenario 2]
    {
        \includegraphics[width=0.8\linewidth]{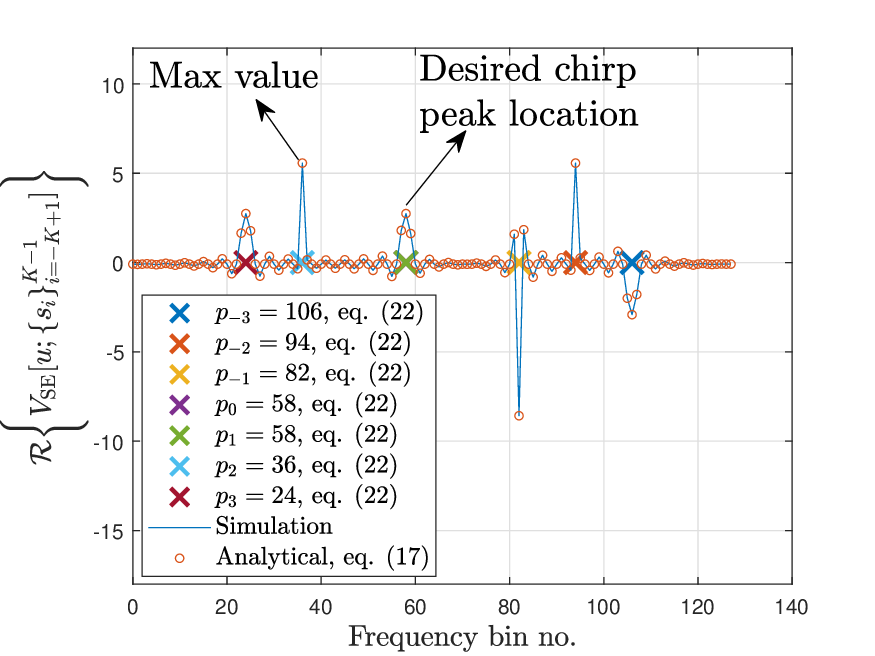}
        
    }

\caption{Illustration of error scenarios 1 and 2 for SE-LoRa with $K=4$ and ${\rm SF}=7$ with the LoRa symbol sequences of (a) $s_{-3}=10$, $s_{-2}=30$, $s_{-1}=20$, $s_{0}=70$, $s_{1}=84$, $s_{2}=100$, and $s_{3}=120$, and (b) $s_{-3}=10$, $s_{-2}=30$, $s_{-1}=50$, $s_{0}=58$, $s_{1}=90$, $s_{2}=100$, and $s_{3}=120$}
  \label{error}
\end{figure}

The work of \cite{superlora} exploits the conventional LoRa detection as in (\ref{LoRadet}) for decoding the SE-LoRa signal. This is equivalent to find the frequency bin corresponding to the maximum value of the real part of $V_{\rm SE}[u;\left\{s_i\right\}_{i=-K+1}^{K-1}]$. However, as can be interpreted from Fig. \ref{reVSE} and Fig. \ref{reVSEK4}, this approach is not effective because the truncated neighboring LoRa chirps introduce interference that manifests as multiple peaks around or on the desired chirp peak location. Even under ideal channel condition ($h=1$ and $w[n]=0$ for $n=0,1,\dots,M-1$), the random generation of LoRa symbol at the transmitter and consequently, the random locations of the peaks $\{p_{i}\}_{i={-K+1}}^{K+1}$, can lead to overlaps. This can cause the loss of the desired chirp peak as the maximum of $\mathcal{R}\{V_{\rm SE}[u;\left\{s_i\right\}_{i=-K+1}^{K-1}]\}$ leading to an error in detection. To further elaborate on this, we introduce two error events resulting from these overlaps after performing conventional LoRa detection on the SE-LoRa scheme:
\begin{enumerate}
    \item Two or more interference positive peaks may be constructively added at the same location, exceeding the desired chirp peak.
    \item One or more interference negative peaks may be destructively added to the desired chirp peak at its location, causing the desired peak to become lower than at least one interference peak.
\end{enumerate}
These error events are illustrated in Fig. \ref{error}. In Fig. \ref{error} (a), the LoRa symbol sequence is selected as $s_{-3}=10$, $s_{-2}=30$, $s_{-1}=20$, $s_{0}=70$, $s_{1}=84$, $s_{2}=100$, and $s_{3}=120$. Exploiting SE-LoRa transmission by this example sequence with $K=4$ results in having two overlapping interference peak locations, i.e., $p_{-1}=p_{1}=52$. The peaks then are constructively added together, and the resulting one becomes higher than that of the desired chirp, i.e., $p_0=70$. This is an example of error event 1. Moreover, in Fig. \ref{error} (b), the example SE-LoRa sequence is selected as $s_{-3}=10$, $s_{-2}=30$, $s_{-1}=50$, $s_{0}=58$, $s_{1}=90$, $s_{2}=100$, and $s_{3}=120$ for $K=4$. It can be observed that one interference peak with a negative value is destructively added to the desired peak chirp at the peak location of $p_1=p_0=58$. As a result, the peak of the desired chirp is no longer the maximum value of the real part of $V_{\rm SE}[u;\left\{s_i\right\}_{i=-K+1}^{K-1}]$, which causes an error in detection due to error event 2.   

To overcome this issue, we propose a SIC-based detection approach based on frequency-domain characteristics of $\mathcal{R}\{V_{\rm SE}[u;\left\{s_i\right\}_{i=-K+1}^{K-1}]\}$. For the detection of a LoRa symbol within the LoRa payload, e.g. $s_q$ ($q=0,\dots,l-1$), the first stage in our proposed SIC-based algorithm is to detect it after canceling the effect of all preceding interference LoRa symbols, i.e. $s_{q-1}$, $s_{q-2}$, $\dots$, and $s_{q-K+1}$, resulting in $\hat{s}_q$, referred to as \textit{first detected version} of $s_{q}$. Afterward, in the second stage, by receiving the upcoming LoRa symbols, up to $s_{q+K-1}$, we refine the detection of $s_q$ by canceling the effect of the succeeding interfering LoRa symbols, resulting in $\doublehat{s}_q$, referred to as the \textit{refined detected version} of $s_q$. Using this approach, we can improve the detection performance. This improvement arises from the fact that the locations of the interfering peaks vary across different receiving windows. Hence, the total probability of error events 1 and 2 becomes very small. In the following, a detailed discussion on the proposed SIC-based detection scheme is provided.

According to the LoRa frame payload structure introduced in Section~IV, we represent the received signal vector $\mathbf{r}_q=[r_q[0],r_{q},\dots,r_q[M-1]]^\top$ as the sampled received signal within the $q$th receiving window ($q=0,1,\dots,l-1$). Consider the first receiving window, corresponding to the desired chirp modulating $s_0$, preceding interfering chirps modulating $\{s_{i}\}_{i=-K+1}^{-1}$, and the succeeding interfering chirps modulating $\{s_{i}\}_{i=1}^{K}$. Note that $\{s_{i}\}_{i=-K+1}^{-1}$ are assumed to be known at the receiver. However, even if the receiver has no information on $\{s_{i}\}_{i=-K+1}^{-1}$, they can be detected using our SIC-based algorithm, similar to the LoRa symbols of the payload part. We can calculate the preceding interference at the first receiving window as
\begin{equation}
\label{r_pre0}
r_{\rm pre,0}[n]=\sqrt{P}h\sum_{i=-K+1}^{-1}x_{\rm L}[n-i\lambda;s_i]g_{M}[n].
\end{equation}
Then, we cancel the effect of $r_{\rm pre,0}[n]$ from the received signal $r_0[n]$ as $r_{\rm pre-c, 0}[n]=r_0[n]-r_{\rm pre,0}[n]$.

By performing conventional LoRa detection on $r_{\rm pre-c, 0}[n]$ as:
\begin{equation}
\label{LoRaDet0}
\hat{s}_0=\argmax_{u\in\mathcal{S}}\mathcal{R}\left\{h^*\times \underbrace{{\rm DFT}\left\{r_{\rm pre-c, 0}[n]x_{\rm L}^*[n;0]\right\}}_{Y_{\rm pre-c, 0}[u]}\right\},
\end{equation}
we can obtain the first detected version of $s_0$ as $\hat{s}_0$.

The next step is to move to the next receiving window (from $\tau$ to $T+\tau$) corresponding to the received signal $r_{1}[n]$ and cancel the effect of the preceding interference by first calculating the preceding interference signal as
\begin{equation}
\label{r_pre1}
r_{\rm pre,1}[n]=\sqrt{P}h\sum_{i=-K+2}^{0}x_{\rm L}[n-i\lambda;s_i]g_{M}[n],
\end{equation}
where $s_i=\hat{s}_i$ for $i\geq0$. Then, we perform conventional LoRa detection on $r_{\rm pre-c, 1}[n]=r_1[n]-r_{\rm pre,1}[n]$ as:
\begin{equation}
\label{LoRaDet1}
\hat{s}_1=\argmax_{u\in\mathcal{S}}\mathcal{R}\left\{h^*\times \underbrace{{\rm DFT}\left\{r_{\rm pre-c, 1}[n]x_{\rm L}^*[n;0]\right\}}_{Y_{\rm pre-c, 1}[u]}\right\},
\end{equation}
and obtain the first detected version of $s_1$ as $\hat{s}_1$. We continue this process until we have a first detected version for every $\{s_i\}_{i=1}^{K-1}$ as $\{\hat{s}_i\}_{i=1}^{K-1}$.

The final step in detecting $s_0$ using our proposed SIC-based detection is to go back and refine the first detected version $\hat{s}_0$ by calculating the succeeding interference signal within the first receiving window using $\{\hat{s}_i\}_{i=1}^{K-1}$ as:
\begin{equation}
\label{r_suc}
r_{\rm suc,0}[n]=\sqrt{P}h\sum_{i=1}^{K-1}x_{\rm L}[n-i\lambda;\hat{s}_i]g_{M}[n].
\end{equation}
Then, by canceling the effect of the succeeding interfering LoRa chirps from $r_{{\rm pre-c},0}[n]$ as $r_{{\rm tot-sic}, 0}[n]=r_{{\rm pre-c},0}[n]-r_{{\rm suc},0}[n]$ and performing LoRa detection as:
\begin{equation}
\label{LoRaDets0final}
\doublehat{s}_0=\argmax_{u\in\mathcal{S}}\mathcal{R}\left\{h^*\times \underbrace{{\rm DFT}\left\{r_{{\rm tot-sic}, 0}[n]x_{\rm L}^*[n;0]\right\}}_{Y_{{\rm tot-sic}, 0}[u]}\right\},
\end{equation}
we obtain the refined detected version of $s_0$ as $\doublehat{s}_0$. 

The next step is to obtain the refined detected version of $s_1$. To this end, first, we calculate the preceding interference for the $K$th receiving window (from $(K-1)\tau$ to $T+(K-1)\tau$) based on the first detected versions $\hat{s}_1$, $\hat{s}_{2}$, $\dots$, and $\hat{s}_{K-1}$ as follows:
\begin{equation}
\label{r_suc4}
r_{{\rm pre},K}[n]=\sqrt{P}h\sum_{i=1}^{K-1}x_{\rm L}[n-i\lambda;\hat{s}_i]g_{M}[n].
\end{equation}
Then, we cancel the effect of this preceding interference from $r_{K}[n]$ as $r_{{\rm pre-c},K}[n]=r_{K}[n]-r_{{\rm pre},K}[n]$. By performing the conventional LoRa detection on $r_{{\rm pre-c},K}[n]$ as:
\begin{equation}
\label{LoRaDetK}
\hat{s}_K=\argmax_{u\in\mathcal{S}}\mathcal{R}\left\{h^*\times \underbrace{{\rm DFT}\left\{r_{{\rm pre-c},K}[n]x_{\rm L}^*[n;0]\right\}}_{Y_{{\rm pre-c},K}[u]}\right\},
\end{equation}
we obtain the first detected version of $s_K$ as $\hat{s}_K$.

Now, we have the estimation of preceding and succeeding interference symbols for refining the detection of $s_1$, i.e., $\{s_{-K+2},s_{-K+3},\dots,s_{-1},\doublehat{s}_0\}$ and $\{\hat{s}_{2},\hat{s}_{3},\dots,\hat{s}_{K}\}$, respectively. Hence, we can calculate the refined preceding interference and succeeding interference signals as:
\begin{equation}
\label{r_preref1}
r_{{\rm pre-ref},1}[n]=\sqrt{P}h\sum_{i=-K+2}^{0}x_{\rm L}[n-i\lambda;s_i]g_{M}[n],
\end{equation}
where $s_i=\doublehat{s}_i$ for $i\geq0$ and 
\begin{equation}
\label{r_suc1}
r_{{\rm suc},1}[n]=\sqrt{P}h\sum_{i=2}^{K}x_{\rm L}[n-i\lambda;\hat{s}_i]g_{M}[n].
\end{equation}
Consequently, the refined detected version of $s_1$ can be obtained as $\doublehat{s}_1$ by performing the conventional LoRa detection on the signal $r_{{\rm tot-sic},1}[n]=r_1[n]-r_{{\rm pre-ref},1}[n]-r_{{\rm suc},1}[n]$ as:
\begin{equation}
\label{LoRaDetref1}
\doublehat{s}_1=\argmax_{u\in\mathcal{S}}\mathcal{R}\left\{h^*\times \underbrace{{\rm DFT}\left\{r_{{\rm tot-sic},1}[n]x_{\rm L}^*[n;0]\right\}}_{Y_{{\rm tot-sic},1}[u]}\right\}.
\end{equation}

This process will be continued until obtaining the refined detected version of every LoRa symbol in the LoRa frame payload. 

\begin{figure*}[t]
  \centering
\includegraphics[width=\linewidth]{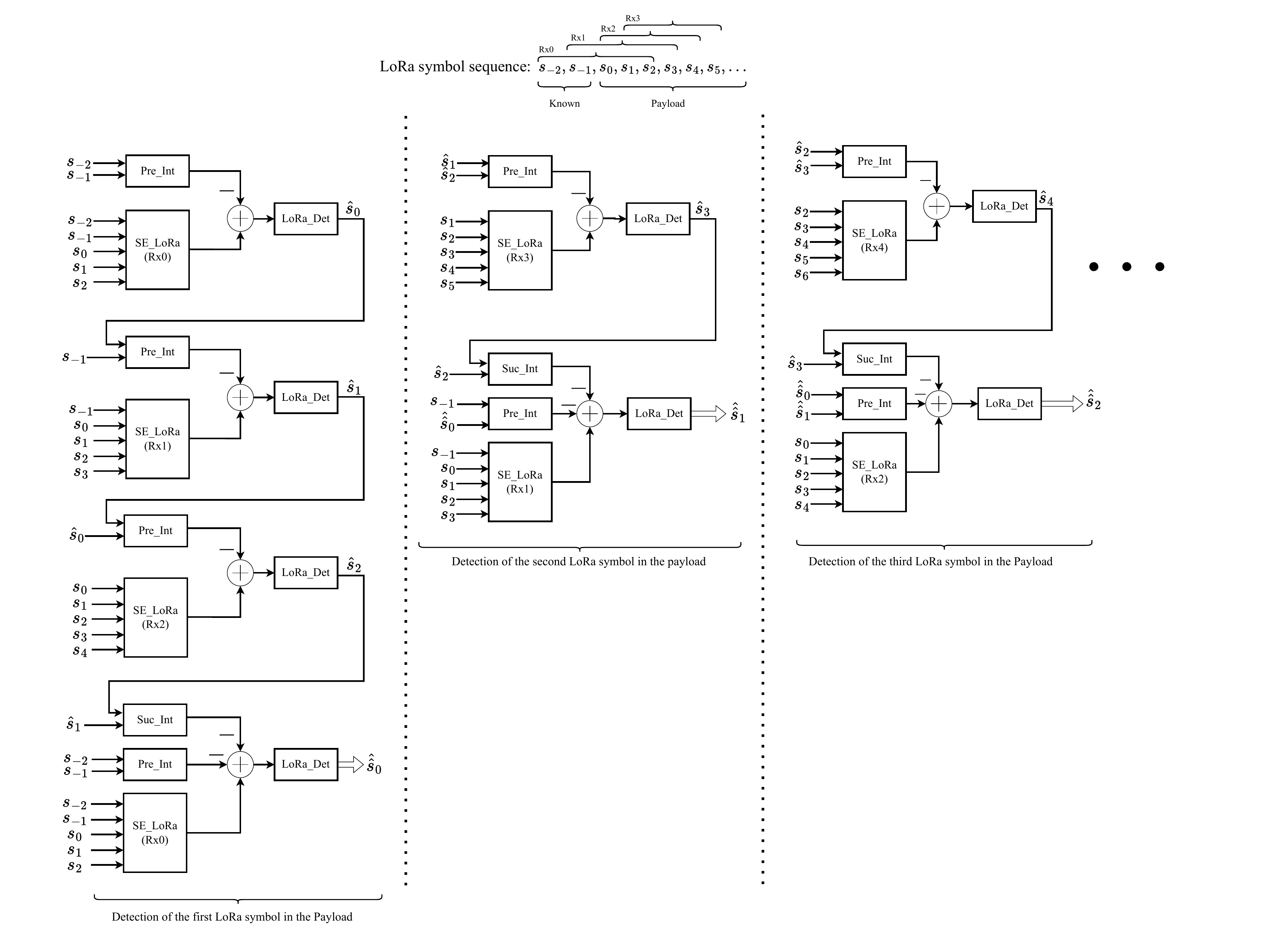}
  \caption{Simplified pseudo-code and flow representation of the proposed low complexity SIC-based detector for $K=3$.}
  \label{psudo}
\end{figure*}

To summarize the proposed SIC-based detector, Algorithm \ref{alg1} is presented at the end of the paper. Moreover, to further illustrate the operations within the Algorithm \ref{alg1}, Fig. \ref{psudo} is presented as a simplified pseudo-code and flow representation of the proposed low complexity SIC-based detector. The blocks in Fig. \ref{psudo} are defined as follows:
\begin{itemize}
    \item Suc\_Int: This block generates a succeeding interference signal which is the summation of $K-1$ truncated LoRa signals. Note that in the signal generation, this block is assumed to have knowledge of $h$ to effectively cancel the interference. 
    \item SE\_LoRa (Rx$i$): This block generates the total received signal in the receiving window of Rx$i$. The output of this block is the summation of all LoRa signals (desired chirp, preceding interference, and succeeding interference) multiplied by the channel gain $h$ and summed by the AWGN noise $w[n]$.
    \item LoRa\_Det: This block performs the conventional LoRa coherent detection on a signal with $M$ samples as expressed in (\ref{LoRadet}).
    \item Pre\_Int: This block generates a preceding interference signal which is the summation of $K-1$ truncated LoRa signals. Note that in the signal generation, this block is assumed to have knowledge of $h$ to effectively cancel the interference.   
\end{itemize}

\begin{algorithm*}[t]
\caption{Proposed low complexity SIC-based SE-LoRa detection}\label{alg1}
\begin{algorithmic}[1]
 \renewcommand{\algorithmicrequire}{\textbf{Input:}}
 \renewcommand{\algorithmicensure}{\textbf{Output:}}
 \REQUIRE $\sqrt{P}$, $h$, $\lambda$, and $r_{\rm win\_ind}[n]$ (${\rm win\_ind}=0,1,\dots,l-1$)
 \STATE For the sake of notation simplicity, here in this algorithm, we define two vectors, i.e., $\mathbf{z}=[z[-K+1],\dots,z[-1]]^\top$ and $\mathbf{c}=[c[1],\dots,c[K-1]]^\top$ as the preceding interfering and succeeding interfering LoRa symbols, respectively.
 \STATE Create the initial preceding symbols vector $\mathbf{z} =[s_{-K+1},\dots,s_{-1}]^\top$ (Note that these $K-1$ symbols are known at the receiver since they are part of the parts prior to the payload)
 \FOR{$q=0,1,\dots,l-1$}
 \IF{$q=0$}
 \STATE Calculate the preceding interference signal $r_{{\rm pre},0}[n]$ as in (\ref{r_pre0}).
\STATE Cancel the the preceding interference from the $0$th received signal to obtain $r_{\rm pre-c, 0}[n]=r_0[n]-r_{\rm pre,0}[n]$.
\STATE Obtain the first detected version of the $0$th LoRa symbol, i.e., $\hat{s}_0$, by performing coherent LoRa detection rule on $r_{{\rm pre-c}, 0}[n]$ as in (\ref{LoRaDet0}).
\FOR{$i_1=1,2,\dots,K-1$}
\STATE Update $\mathbf{z}=[\hat{s}_{-K+1+i_1},\dots,\hat{s}_{-1+i_1}]^\top$ (Note that $\hat{s}_{-K+1+i_1}=s_{-K+1+i_1}$ for $i_1<K-1$.
 \STATE Calculate $r_{{\rm pre, temp}}[n]=\sqrt{P}h\sum_{i=-K+1}^{-1}x_{\rm L}[n-i\lambda;z[i]]g_{M}[n]$.
\STATE Cancel the the preceding interference from the received signal as $r_{\rm pre-c,temp}[n]=r_{i_1}[n]-r_{\rm pre,temp}[n]$.
\STATE Obtain the first detected version $\hat{s}_{i_1}$, by performing coherent LoRa detection on $r_{{\rm pre,temp}}[n]$ as:
\begin{equation}
\label{LoRaDettemp}
\hat{s}_{i_1}=\argmax_{u\in\mathcal{S}}\mathcal{R}\left\{h^*\times \underbrace{{\rm DFT}\left\{r_{{\rm pre,temp}}[n]x_{\rm L}^*[n;0]\right\}}_{Y_{{\rm pre,temp}}[u]}\right\}.
\end{equation}
\ENDFOR
\STATE Create the succeeding symbols vector $\mathbf{c}=[\hat{s}_{1},\hat{s}_{2},\dots,\hat{s}_{K-1}]$.
\STATE Calculate the succeeding interference signal $r_{{\rm suc},0}[n]$ as in (\ref{r_suc}).
\STATE Cancel the the succeeding interference as $r_{{\rm tot-sic}, 0}[n]=r_{{\rm pre-c},0}[n]-r_{{\rm suc},0}[n]$.
\STATE Obtain the refined detected version of the $0$th LoRa symbol, i.e., $\doublehat{s}_0$, by performing coherent LoRa detection rule on $r_{{\rm tot-sic}, 0}[n]$ as in (\ref{LoRaDets0final}).
\STATE Update $\mathbf{z}=[{s}_{-K+2},\dots,s_{-1},\doublehat{s}_{0}]^\top$.
\ELSE
\STATE Calculate the preceding interference signal for $\mathbf{r}_{K+q-1}$ as $r_{{\rm pre},K+q-1}[n]=\sqrt{P}h\sum_{i=1}^{K-1}x_{\rm L}[n-i\lambda;c[i]]g_{M}[n]$.
\STATE Cancel the the preceding interference from the received signal $\mathbf{r}_{K+q-1}$ as $r_{{\rm pre-c},K+q-1}[n]=r_{K+q-1}[n]-r_{{\rm pre},K+q-1}[n]$.
\STATE Obtain the first estimation of LoRa symbol, i.e., $\hat{s}_{q+K-1}$, by performing coherent LoRa detection rule on $r_{{\rm pre-c}, q+K-1}[n]$ as:
\begin{equation}
\label{LoRaDetqk}
\hat{s}_{q+K-1}=\argmax_{u\in\mathcal{S}}\mathcal{R}\left\{h^*\times \underbrace{{\rm DFT}\left\{r_{{\rm pre-c}, q+K-1}[n]x_{\rm L}^*[n;0]\right\}}_{Y_{{\rm pre-c}, q+K-1}[u]}\right\}.
\end{equation}
\STATE Update $\mathbf{c}=[\hat{s}_{1+q},\hat{s}_{2+q},\dots,\hat{s}_{K-1+q}]^\top$
\STATE Calculate the refined preceding interference signal for $q$th receiving window as $r_{{\rm pre-ref},q}[n]=\sqrt{P}h\sum_{i=-K+1}^{-1}x_{\rm L}[n-i\lambda;z[i]]g_{M}[n]$.
\STATE Calculate the succeeding interference signal for $q$th receiving window as $r_{{\rm suc},q}[n]=\sqrt{P}h\sum_{i=1}^{K-1}x_{\rm L}[n-i\lambda;c[i]]g_{M}[n]$.
\STATE Cancel the the preceding and succeeding interference from the received signal $r_q[n]$ as $r_{{\rm tot-sic},q}[n]=r_q[n]-r_{{\rm pre-ref},q}[n]-r_{{\rm suc},q}[n]$.
\STATE Obtain the refined detected version $\doublehat{s}_{q}$, by performing coherent LoRa detection rule on $r_{{\rm tot-sic}, q}[n]$ as:
\begin{equation}
\label{LoRaDetq}
\doublehat{s}_{q}=\argmax_{u\in\mathcal{S}}\mathcal{R}\left\{h^*\times \underbrace{{\rm DFT}\left\{r_{{\rm tot-sic}, q}[n]x_{\rm L}^*[n;0]\right\}}_{Y_{{\rm tot-sic}, q}[u]}\right\}.
\end{equation}
\STATE Update $\mathbf{z}=[\doublehat{s}_{-K+2+q},\dots,\doublehat{s}_{q}]^\top$ (Note that $\doublehat{s}_{-K+2+q}={s}_{-K+2+q}$ for $q<K-2$).
\ENDIF
\ENDFOR

 \end{algorithmic} 
 \end{algorithm*}

It should be noted that the computational complexity of the proposed low complexity SIC-based detector is $\mathcal{O}(M+K)$, as can be seen from Fig. \ref{psudo}. This is much lower than that of joint ML detection with a complexity of $\mathcal{O}(lM^{l+1})$.

\section{Simulation Results}
In this section, we evaluate the performance of the proposed low complexity SIC-based detector for SE-LoRa modulation in terms of SER. It should be noted that the simulation parameters are selected according to LoRaWAN specification in the U.S. region \cite{9} as presented in Table \ref{simpar}. Moreover, for environmental modeling, i.e., channel fading, we consider three cases:
\begin{itemize}
    \item Ideal environment with a dominant line-of-sight (LoS) link between the LoRa EDs and the LoRa GW, and no fading is modeled as AWGN. 
    \item Urban areas with no LoS links between the LoRa EDs and the LoRa GW is modeled as Rayleigh fading along with AWGN.
    \item Rural areas with both LoS and non-LoS links between the LoRa EDs and the LoRa GW is modeled as Rician fading (with Rician factor of $6$ dB) along with AWGN \cite{Rician}.
\end{itemize}
\begin{table}[t!]
\caption{Simulation parameters \cite{9}.}
\label{simpar}
\centering
\begin{tabularx}{\linewidth}{Xc}
\toprule[1.0pt]
LoRa ED Tx power & $15$ dBm \\
LoRa payload length, SF7 & $57$ LoRa symbols \\
$B$, SF7 & $125$ kHz \\
LoRa payload length, SF9 & $44$ LoRa symbols \\
$B$, SF9 & $125$ kHz \\
LoRa payload length, SF11 & $36$ LoRa symbols \\
$B$, SF11 & $500$ kHz \\
\bottomrule[1.0pt]
\end{tabularx}
\end{table}

It also should be noted that the term ``Conv. Det.'' in the legend of simulation result figures refers to detecting SE-LoRa signal using conventional LoRa detection, as provided in \cite{superlora}.

\begin{figure}[t]
  \centering
\includegraphics[width=0.9\linewidth]{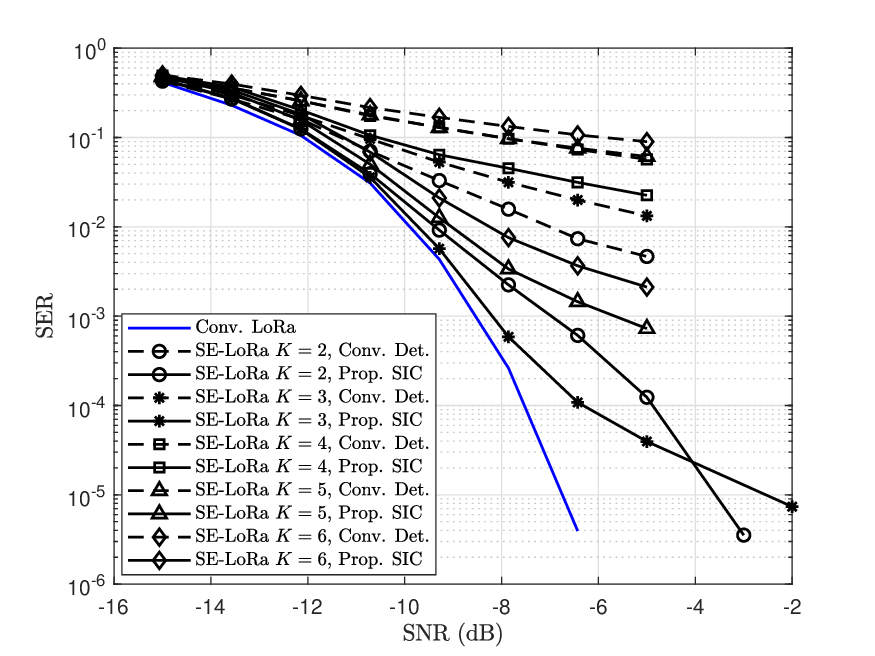}
  \caption{SER for AWGN channel with ${\rm SF}=7$, $B=125$ kHz, and payload length of $57$ LoRa symbols.}
  \label{AWGN_Prop}
\end{figure}

Fig. \ref{AWGN_Prop} shows the SER performance of the SE-LoRa modulation in AGWN channel condition with ${\rm SF}=7$, $B=125$ kHz, and payload length of $57$ LoRa symbols. As illustrated, the proposed low complexity SIC-based detection algorithm outperforms the conventional detector. However, since LoRa modulation inherently performs well under low-SNR conditions in AWGN channels, the low complexity SIC-based detector experiences a considerable performance gap compared to the conventional LoRa modulation. This occurs because, during the interference cancellation stage, the noise power becomes comparable to the interference peaks observed at the output of the DFT of the dechirped received signal, which may lead to imperfect cancellation of those interference components. 

\begin{figure}[t]
  \centering
\includegraphics[width=0.9\linewidth]{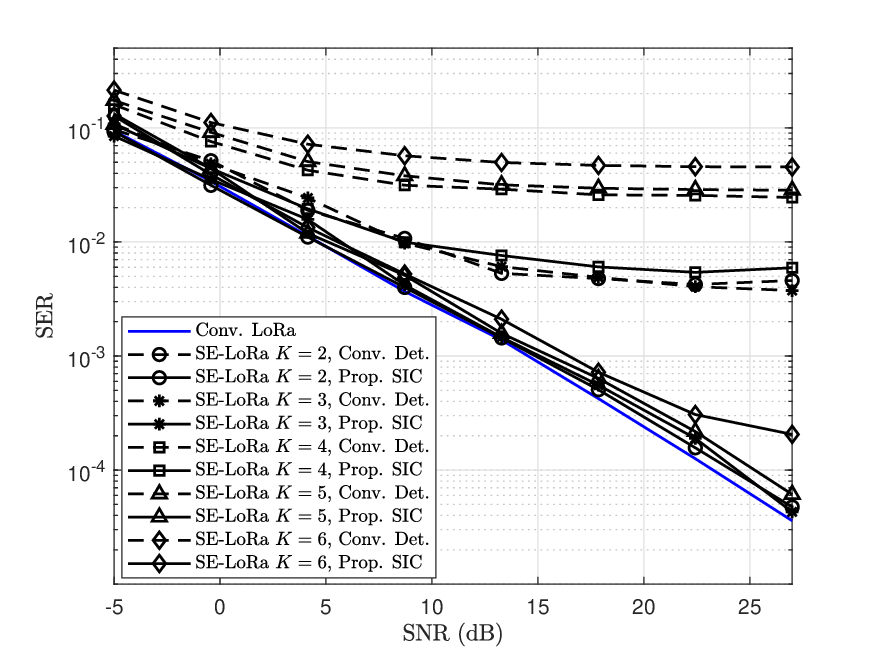}
  \caption{SER for Rayleigh channel with ${\rm SF}=7$, $B=125$ kHz, and payload length of $57$ LoRa symbols.}
  \label{Rayleigh_Prop}
\end{figure}

Fig. \ref{Rayleigh_Prop} illustrates the SER performance of the low complexity SIC-based detector in the Rayleigh channel condition with ${\rm SF}=7$, $B=125$ kHz, and payload length of $57$ LoRa symbols. It can be observed that, since LoRa modulation requires relatively high SNRs to achieve acceptable performance under Rayleigh fading, the noise level is no longer comparable to the interference peaks as it was in the AWGN case. Consequently, the proposed algorithm can effectively cancel interference. As shown, even for $K=5$, the performance gap between the conventional LoRa modulation (without interference) and the proposed SE-LoRa modulation with the low complexity SIC-based detector is around $1$ dB, which is acceptable. 

\begin{figure}[t]
  \centering
\includegraphics[width=0.9\linewidth]{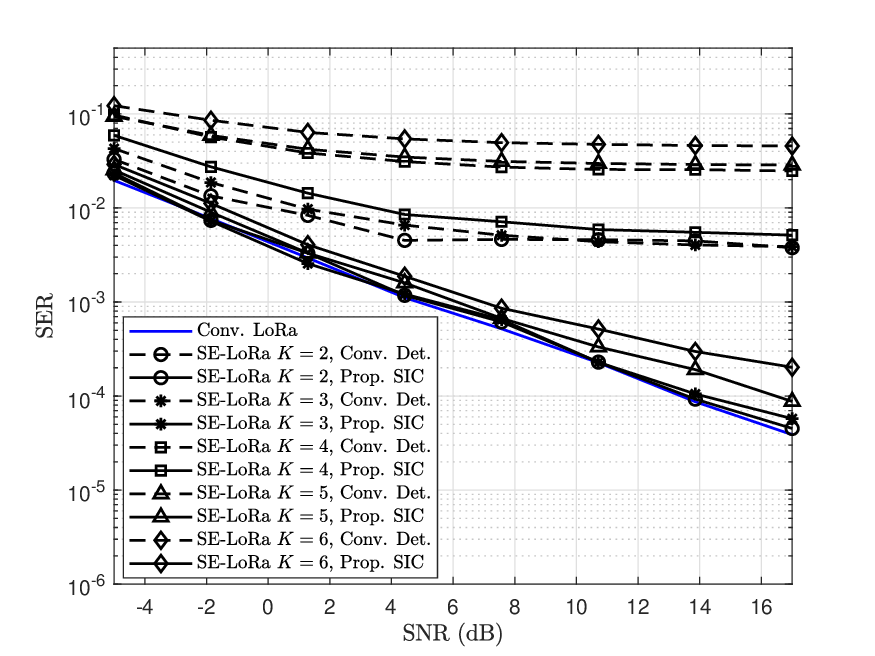}
  \caption{SER for Rician channel with ${\rm SF}=7$, $B=125$ kHz, and payload length of $57$ LoRa symbols.}
  \label{Rician_SF7}
\end{figure}

\begin{figure*}[t]
  \centering
\includegraphics[width=0.8\linewidth]{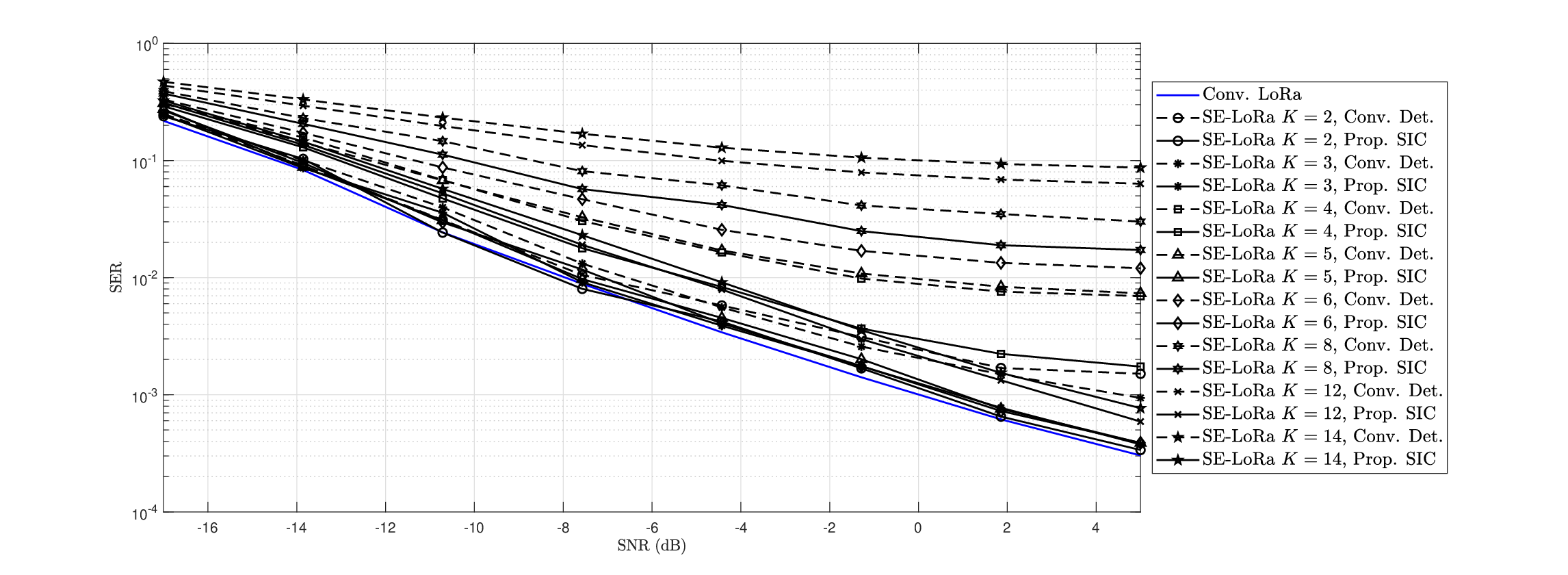}
  \caption{SER for Rician channel with ${\rm SF}=9$, $B=125$ kHz, and payload length of $44$ LoRa symbols.}
  \label{Rician_SF9}
\end{figure*}

\begin{figure*}[t]
  \centering
\includegraphics[width=0.8\linewidth]{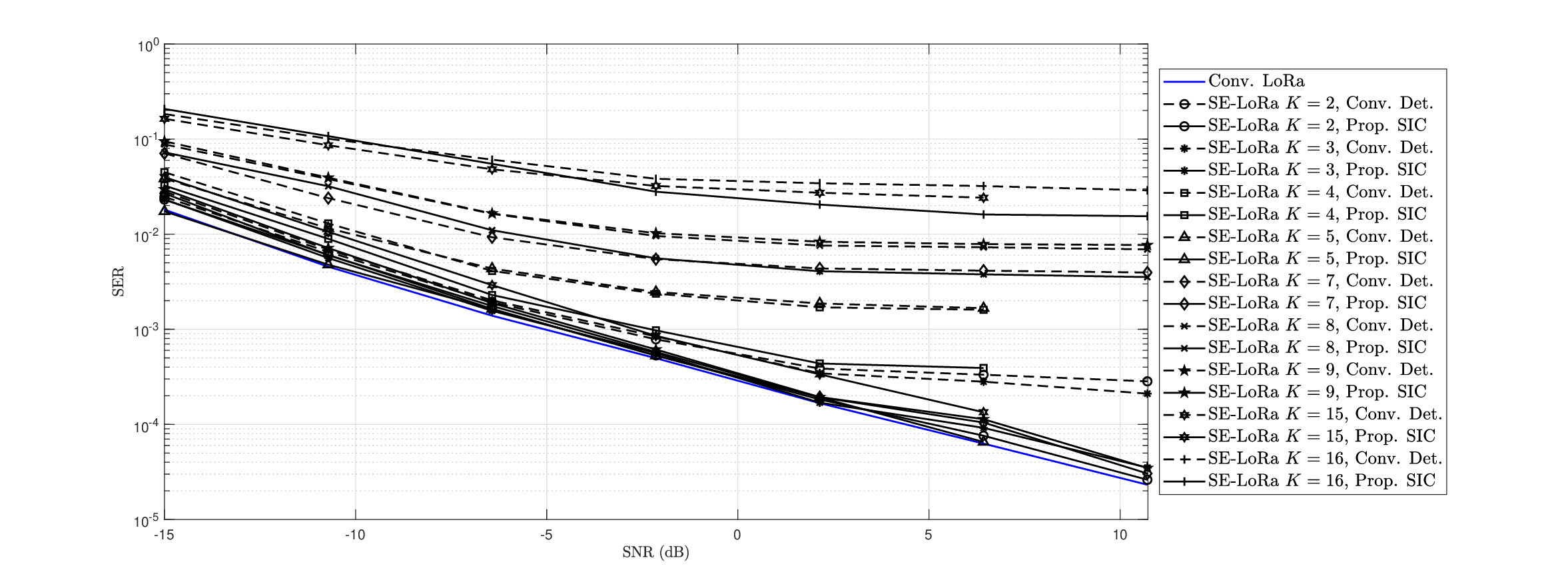}
  \caption{SER for Rician channel with ${\rm SF}=11$, $B=500$ kHz, and payload length of $32$ LoRa symbols.}
  \label{Rician_SF11}
\end{figure*}

In the following, Fig. \ref{Rician_SF7} (with ${\rm SF}=7$, $B=125$ kHz, and payload length of $57$ LoRa symbols), Fig. \ref{Rician_SF9} (with ${\rm SF}=9$, $B=125$ kHz, and payload length of $44$ LoRa symbols), and Fig. \ref{Rician_SF11} (with ${\rm SF}=11$, $B=500$ kHz, and payload length of $32$ LoRa symbols) are presented to evaluate the performance of the low complexity SIC-based detector in Rician fading channel (representing rural areas with Rician factor equal
to $6$ dB \cite{Rician}).  As observed, the low complexity SIC-based detector effectively mitigates interference in the SE-LoRa transmission scheme for most values of $K$ and achieve an acceptable performance compared to the case of conventional LoRa modulation without the interference. 

Considering Fig. \ref{Rician_SF7}, Fig. \ref{Rician_SF9}, and Fig. \ref{Rician_SF11}, as discussed in Section~III, increasing the SF reduces the data rate in LoRa modulation, which eventually improves the error performance \cite{Maleki_chirp}. However, a larger SF also increases $M$, which expands the number of frequency bins. This increase in the number of frequency bins, along with the higher bandwidth requirements for the case of ${\rm SF}=11$, results in a lower probability of overlapping between the desired chirp and interference peaks at the DFT and dechirping operations. Hence, the interference can be canceled more effectively because the error events 1 and 2 (as discussed in Section~V) are less likely to happen. Consequently, our proposed low complexity SIC-based detector can perform well for $K$ values up to $15$ for the case of ${\rm SF}=11$ and $B=500$ kHz. This makes our proposed SIC-based detector a very well-suited solution for implementing a reliable LoRa-based network for applications demanding higher data rates.

\begin{figure}[t]
  \centering
\includegraphics[width=0.9\linewidth]{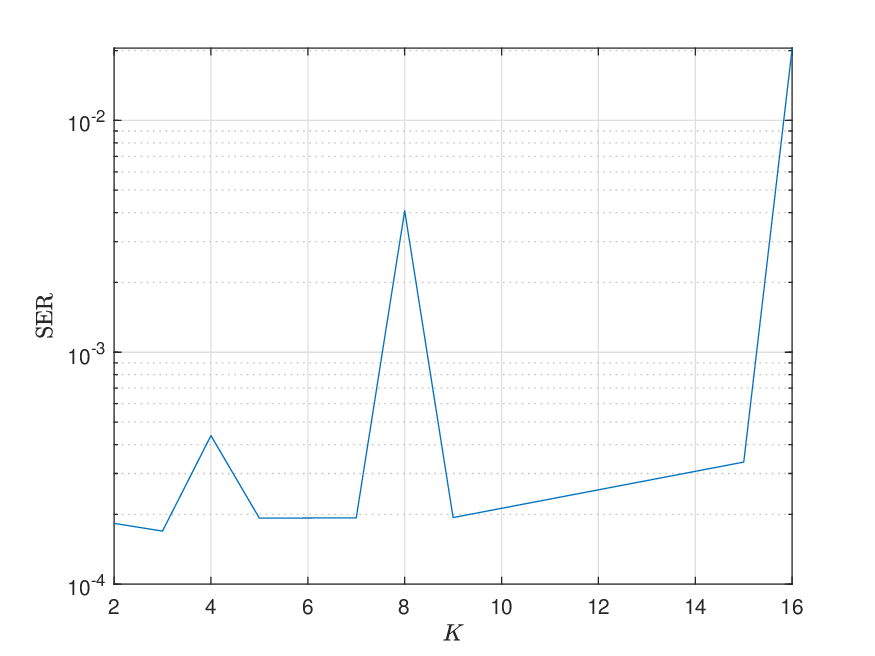}
  \caption{SER performance of low complexity SIC-based detector for different values of $K$ at SNR of $2.1$ dB, ${\rm SF}=11$, and Rician channel condition.}
  \label{Ptwo}
\end{figure}

\begin{figure*}[t]
  \centering
\subfigure[$K=3$]
    {
        \includegraphics[width=\linewidth]{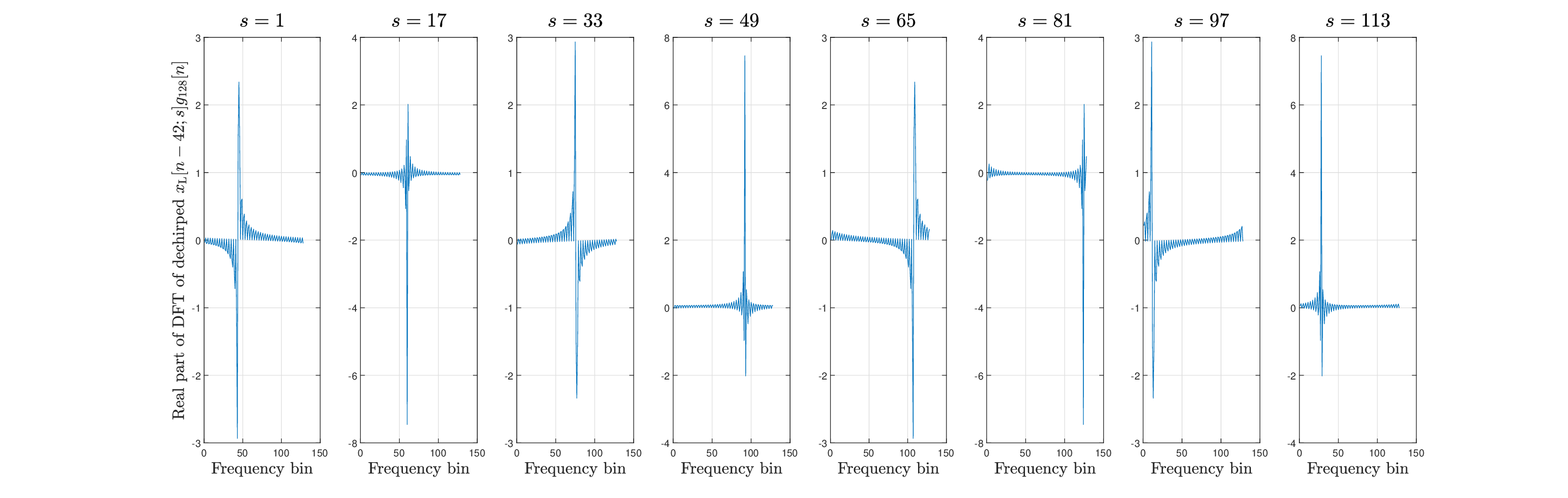}
        
    }
\\
\subfigure[$K=4$]
    {
        \includegraphics[width=\linewidth]{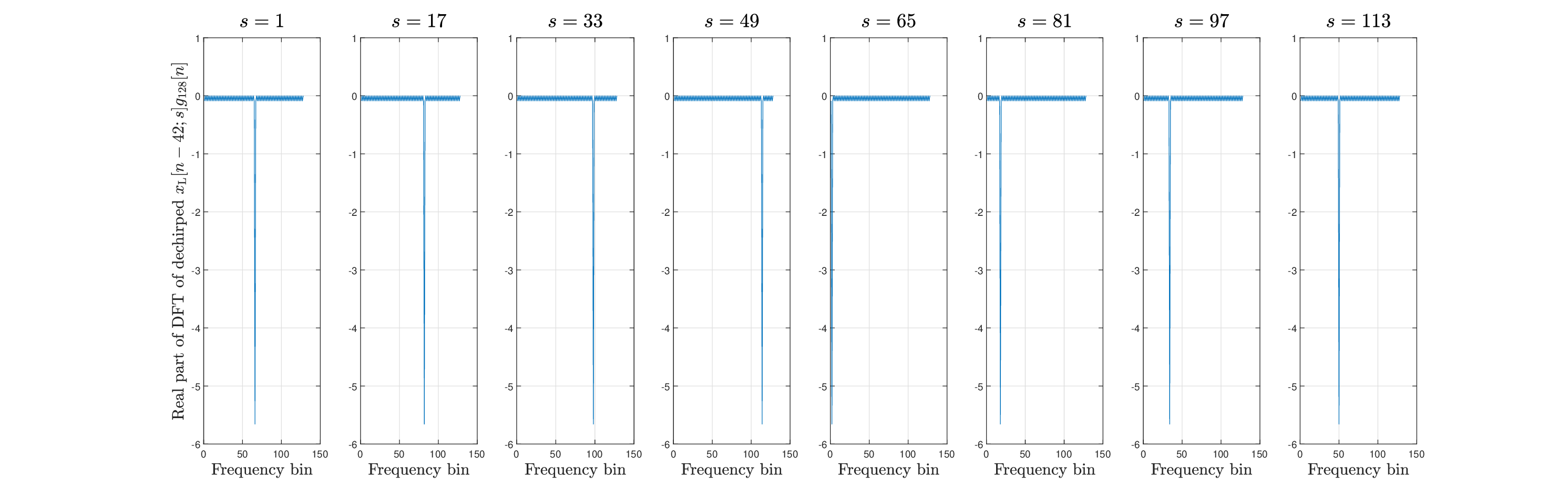}
        
    }

\caption{Illustration of frequency-domain behavior of dechirped of second succeeding interfering LoRa signal $x_{\rm L}[n-i\lambda;s]g_{M}[n]$ ($i=2$) for $M=128$ and various values for $s$.}
  \label{just}
\end{figure*}

Finally, as can be noticed from the provided SER results, an irregular behavior can be observed in the SER performance for $K=4$, $K=8$, and $K=16$ compared to other $K$ values. Fig. \ref{Ptwo} is a clear illustration of this behavior at SNR of $2.1$ dB. 
This anomaly arises because, when $K$ is a power of two, the interference peaks appear as distinct, sharp positive or negative peaks (depending on the value of LoRa symbol) for all LoRa symbols. This symmetry can lead to errors in the interference cancellation stage, where the detector may inadvertently suppress the desired chirp peak or amplify adjacent interference components. In contrast, when $K$ is not a power of two, the interference peaks typically exhibit both positive and negative parts around the peak location, reducing the likelihood of such cancellation errors. As an example, Fig. \ref{just} is presented as a comparison between (a) $K=3$ and (b) $K=4$. The results provided in Fig. \ref{just} (a) are obtained as the real part of DFT of the dechirped truncated LoRa signal $x_{\rm L}[n-i\lambda;s]g_{M}[n]$ for $M=128$ and $i=2$, indicating the second succeeding interfering LoRa signal. Similarly, for the case of $K=4$, the same results are provided in Fig. \ref{just} (b). It can be seen that for the case of $K=4$, the interfering signal shows a sharp negative peak regardless of the value of $s$, while the interfering peak for $K=3$ exhibits both negative and positive parts around the peak location. 

To conclude the simulation results, Table \ref{wrap} (at the end of the paper) summarizes the SER performance of SE-LoRa modulation with the low complexity SIC-based detection, highlighting both the performance gap relative to conventional LoRa modulation without interference and the spectral efficiency gain achieved through faster transmission of LoRa signals. It should be noted that there is a trade-off between the error performance and the spectral efficiency improvement of SE-LoRa settings and one can choose the value of $K$ accordingly, considering the two parameters presented in Table \ref{wrap}, i.e., $G_{\rm SE}$ and $L_{\rm SER}$. The parameter $G_{\rm SE}$ is the spectral efficiency improvement percentage compared to the conventional LoRa, calculated as $(\eta_{\rm SE-L}-\eta_{\rm L})/\eta_{\rm L} \times 100\%$ according to (\ref{SELoRa}) and (\ref{SESELoRa}). Moreover, $L_{\rm SER}$ is the difference between the required SNR for obtaining a SER of $10^{-3}$ for conventional LoRa and SE-LoRa. Based on the results in Table \ref{wrap}, it can be concluded that SE-LoRa modulation with the low complexity SIC-based detection provides a highly desirable improvement in spectral efficiency while maintaining an error performance that is fully comparable to that of conventional LoRa modulation. 

\begin{table}[t]
\caption{Comparison between SE-LoRa modulation with the low complexity SIC-based detector and conventional LoRa modulation in terms of SER and spectral efficiency (for a payload of $50$ LoRa symbols, i.e., $l=50$)}
\label{wrap}
\centering
\begin{tabularx}{0.7\linewidth}{cccc}
\toprule[0.5pt]
\textbf{Channel and SF} & \textbf{$K$} & $G_{\rm SE}$ & $L_{\rm SER}$ (dB) \\ 
\midrule[0.5pt] 
\vspace{5pt}
\multirow{3}{*}{AWGN, ${\rm SF}=7$}& $2$ & $96.08\%$ & $1.5$ \\
\vspace{5pt}
 & $3$ & $188.46\%$ & $<0.5$ \\
\vspace{5pt}
 & $5$ & $362.96\%$ & $2.8$ \\
\midrule[0.5pt]
\vspace{5pt}
\multirow{4}{*}{Rayleigh, ${\rm SF}=7$}& $2$ & $96.08\%$ & $<0.5$ \\
\vspace{5pt}
 & $3$ & $188.46\%$ & $<0.5$ \\
\vspace{5pt}
 & $5$ & $362.96\%$ & $1$ \\
 \vspace{5pt}
 & $6$ & $445.45\%$ & $2.5$ \\
 \midrule[0.5pt]
\vspace{5pt}
\multirow{4}{*}{Rician, ${\rm SF}=7$}& $2$ & $96.08\%$ & $<0.5$ \\
\vspace{5pt}
 & $3$ & $188.46\%$ & $<0.5$ \\
\vspace{5pt}
 & $5$ & $362.96\%$ & $1.5$ \\
 \vspace{5pt}
 & $6$ & $445.45\%$ & $2.5$ \\
 \midrule[0.5pt]
\vspace{5pt}
\multirow{4}{*}{Rician, ${\rm SF}=9$}& $2$ & $96.08\%$ & $<0.5$ \\
\vspace{5pt}
 & $3$ & $188.46\%$ & $<0.5$ \\
\vspace{5pt}
 & $5$ & $362.96\%$ & $1$ \\
 \vspace{5pt}
 & $6$ & $445.45\%$ & $1$ \\
  \vspace{5pt}
 & $12$ & $883.61\%$ & $2.5$ \\
  \vspace{5pt}
 & $14$ & $1011.11\%$ & $3$ \\
\midrule[0.5pt]
\vspace{5pt}
\multirow{4}{*}{Rician, ${\rm SF}=11$}& $2$ & $96.08\%$ & $<0.5$ \\
\vspace{5pt}
 & $3$ & $188.46\%$ & $1$ \\
\vspace{5pt}
 & $5$ & $362.96\%$ & $1.5$ \\
 \vspace{5pt}
 & $7$ & $525\%$ & $1.5$ \\
  \vspace{5pt}
 & $9$ & $675.86\%$ & $2$ \\
  \vspace{5pt}
 & $15$ & $1071.88\%$ & $2.5$ \\
\bottomrule[0.5pt]
\end{tabularx}
\end{table}

\section{Conclusion}
In this paper, we proposed a new LoRa-based modulation scheme, i.e., SE-LoRa, with low complexity SIC-based detector. After formulating the joint ML detection, to tackle its very high computational complexity, we propose a low complexity detector inspired by the frequency-domain characteristics of dechirped SE-LoRa modulation. Our detector uses SIC to effectively cancel the effect of interfering LoRa signals resulted from faster transmission and results in a desirable error performance compared to the conventional LoRa modulation. Using computer simulations, it is observed that exploiting SE-LoRa with our proposed SIC-based detection in Rician channels, for spectral efficiency improvement up to $445.45\%$, $1011.11\%$, and $1071.88\%$ for SF values of $7$, $9$, and $11$, respectively, we only lose less than $3$ dB in performance at SER of $10^{-3}$. This spectral efficiency improvement along with the acceptable error performance of SE-LoRa with low complexity SIC-based detector can make it a potential candidate to be exploited in high data rate LoRaWAN-based IoT applications.

\bibliographystyle{IEEEtran}
\bibliography{IoT_Transmission_Techs_PhD_V20}

\end{document}